\documentclass{aa} 
\usepackage{graphicx}
\usepackage{txfonts}
\usepackage{xcolor}
\begin{document}

   \title{Simulating AIA observations of a flux rope ejection}


\author{P.Pagano\inst{1} \and D.H. Mackay\inst{1} \and S.Poedts\inst{2}}
\institute{{School of Mathematics and Statistics, University of St Andrews, North Haugh, 
St Andrews, Fife, Scotland KY16 9SS, UK.}
\and
{Dept. of Mathematics, Centre for Mathematical Plasma Astrophysics, KU Leuven, Celestijnenlaan 200B, 3001 Leuven, Belgium.}
}

   \date{}

 
  \abstract
   {Coronal mass ejections (CMEs) are the most violent phenomena observed on the Sun. Currently, extreme ultraviolet (EUV) images from the
Atmospheric Imaging Assembly (AIA) on board the Solar Dynamic Observatory (SDO) are providing new insights into the early phase of CME evolution.
In particular, observations now show the ejection of magnetic flux ropes from the solar corona 
and how they evolve into CMEs. While this is the case, these observations are difficult to interpret in terms of
basic physical mechanisms and quantities.
To fully understand CMEs we need to compare equivalent quantities derived
from both observations and theoretical models.
This will aid in bridging the gap between observations and models.}
   {To this end we aim
to produce synthesised AIA observations from simulations of a flux rope ejection.
To carry this out we include the role of thermal conduction and radiative losses,
both of which are important for determining
the temperature distribution of the solar corona during a CME.}
   {We perform a simulation where a flux rope is ejected from the solar corona.
From the density and temperature of the plasma in the simulation
we synthesise AIA observations.
The emission is then integrated along the line of sight using the 
instrumental response function of AIA.}
   {We sythesise observations of AIA in the channels at $304\AA$, $171\AA$, $335\AA$, and $94\AA$.
The synthesised observations show a number of features similar to actual observations
and in particular reproduce the general development of CMEs in the low corona as observed by AIA.
In particular we reproduce an erupting and expanding arcade in the $304\AA$ and $171\AA$ channels
with a high density core.}
   {The ejection of a flux rope reproduces many of the features found in the AIA observations.
This work is therefore a step forward in bridging the gap between observations and models,
and can lead to more direct interpretations of EUV observations in terms of flux rope ejections.
We plan to improve the model in future studies in order to perform a more quantitative comparison.}


   \maketitle
%

\section{Introduction}
Coronal mass ejections (CMEs) are violent eruptions of plasma and magnetic flux from the solar corona.
They have been observed since 1971 \citep{Tousey1973} with increasingly more accurate and precise instruments.
While most of the satellites built to study the Sun investigate CMEs,
many aspects of CMEs remain an enigma and outstanding questions are being actively investigated by the solar community.
One of the most recent satellites to study CMEs is the Solar Dynamic Observatory (SDO).
In particular the Atmospheric Imaging Assembly instrument \citep[AIA;][]{Lemen2012}
continuously observes the full solar disk providing observations in
extreme ultraviolet (EUV) channels.
Images from AIA enable us to observe
with high spatial and temporal resolution
the early phase of ejections and the regions where they originate.
Because of this, the underlying physics of the generator and propagation of CMEs can be studied.
To date, solar observations stimulate many questions that theoretical models are presently not able to answer.
To bridge the gap between observations and models, we must determine
which features our models are already able to describe and what remains to be understood.
This can only be carried out if equivalent quantities
from theoretical models and observations can be directly compared.

Many models have been put forward to explain CMEs.
One such model is the flux rope ejection model, where
a magnetic flux rope is formed in the corona and is subsequently
violently ejected \citep{ForbesIsenberg1991,Amari2000,FanGibson2007}.
Whether the flux rope ejection model can explain all CMEs, or only some of them,
is still an open question.
However direct observations have clearly linked
the ejection of a flux rope with at least 40\% of CMEs \citep{Vourlidas2013}.
For the remaining CMEs we do not have sufficient information to determine
if a flux rope is present or not.
Only in a small percentage of cases can we exclude the presence of a flux rope.
Moreover, additional studies, directly involving AIA, have shown that flux ropes are formed and ejected 
during a CME \citep{Chen2011,LiZhang2013}.
From a theoretical point of view, a number of studies have shown that the ejection of
a flux rope is sufficient to reproduce the main observed features of CMEs:
\citet{TorokKliem2007}, \citet{Fan2010}, and \citet{Aulanier2010} explain the ejection with the occurrence of a Torus instability;
\citet{Fan2009} and \citet{Archontis2009} with a flux emergence event;
\citet{Savcheva2012} with the rotation of footpoints;
\citet{Amari2003} with shearing and flux cancellation, while
\citet{Amari2011} use convergence of foot points;
finally \citet{Roussev2012} consider a global reorganisation of the solar corona.
\citet{Zuccarello2012} used STEREO observations to produce a 3D reconstruction of the early
evolution path of a CME and compared it to numerical simulations
in order to determine the role of streamers in the deflection of CMEs towards the equatorial plane.
Additionally, \citet{Pagano2013a} and \citet{Pagano2013b}
have developed an approach where the flux rope ejection is modelled with magnetohydrodynamics (MHD) simulations.
They start from an already out of equilibrium initial condition
where the flux rope formation generates the Lorentz force excess for the ejection.
A technical innovation of this approach is that
the flux rope formation is described by a global non-linear force-free field (GNLFFF)
model \citep{MackayVanBallegooijen2006A} and the final state of the GNLFFF is coupled with 
the initial condition of the MHD simulation.
We base a significant part of the present work on \citet{Pagano2013a} and \citet{Pagano2013b}
and the approach developed within them.

The aim of this paper is to generate a technique that will enable us to directly
compare the flux rope ejection model of \citet{Pagano2013b}
with actual observations, in particular EUV emission seen by AIA.
Our goal is to develop a tool with which to check the validity of our model and to understand what needs
to be improved to accurately describe solar eruptions.
The strategy we follow is to reproduce AIA observations from a MHD simulation and then
analyse these synthesised observations in order to determine differences and similarities with the actual observations.
In order to pursue our strategy we perform a MHD simulation
with an initial condition based on the parameter study carried out in \citet{Pagano2013b}
where a flux rope is violently ejected.
We also tune the thermodynamic parameters of the simulation to have
a non-isothermal initial profile, which provides
the flux rope with realistic values of density and temperature required for comparison with observations
where it is represented by a structure that is denser and colder compared to its surroundings.
During the evolution of the flux rope we compute the EUV emission from each plasma element of the MHD simulation.
The emission is then integrated along the line of sight
taking into account the instrumental response of each AIA channel.
In this way we reproduce synthetic AIA images from our MHD model that
can be analysed and compared with actual observations.
Our strategy has already been applied by other authors.
\citet{Pagano2008} reproduced UVCS/SoHo observations to study the passage of shocks connected to CMEs in the solar corona.
\citet{Hoilijoki2013} calculated the resulting emission measure from a MHD simulation to study the propagation of MHD waves
and \citet{Roussev2012} reproduced the soft X-ray emission from their simulation.
In other studies \citet{Downs2011} reproduced EUV images from a CME simulation in a global context,
and \citet{Lugaz2011} reproduced an observed flux rope ejection and synthesised the corresponding EIT images.

Another aim of this work is to assess the importance of the effects of thermal conduction and radiative losses when modelling a CME.
Some previous work has already taken important steps towards a realistic 
physical modelling of the solar corona.
For example, \citet{Mok2005} studied the detailed thermal structure of an active region including the transition region, 
\citet{Lionello2009} simulated the global corona over a period of several weeks including non-ideal MHD effects.
Through this realistic model the authors developed an online tool with which
to perform physically realistic EUV syntheses of a corona in near-equilibrium.
Finally, \citet{Xia2014} have simulated the formation of a twisted flux rope in a non-ideal MHD framework.

The present paper intends to put forward a simple but useful technique to study flux rope ejections.
From this, the long term goal is to carry out a comparison of equivalent quantities
found in both EUV observations and EUV synthesised images.
Of course, in order to significantly expand our knowledge of CMEs,
in the near future we must go beyond the results of the present paper.
We set out ambitious goals for future research, such as the study of specific CME events,
or the focus on individual structures observed during flux rope ejections (e.g. current sheets or shocks).

The paper is structured as follows.
In Sect. \ref{mhdsimulation} we describe how the MHD simulation is set up and the evolution of the simulation.
In Sect. \ref{aiasdoemissionsythesis} we describe our technique used to reproduce AIA observations from the MHD simulation.
Following this, in Sect. \ref{results} we show the results of our synthesised observations and the features they produce.
Finally, in Sect. \ref{discussion} we discuss some aspects of the present work and
in Sect. \ref{conclusion} we draw some general conclusions.

\section{MHD Simulation}
\label{mhdsimulation}
In order to reproduce the EUV emission of the coronal plasma during a flux rope ejection,
we first perform a MHD simulation where a flux rope ejection occurs.
Subsequently, from the MHD simulation which provides the plasma density and temperature
as a function of time and position,
we can infer the emission observable by AIA.
We use the MPI-AMRVAC software~\cite{Keppens2012}, to solve the MHD equations
where external gravity, anisotropic thermal conduction, and optically thin radiative losses are included
\begin{equation}
\label{mass}
\frac{\partial\rho}{\partial t}+\vec{\nabla}\cdot(\rho\vec{v})=0,
\end{equation}
\begin{equation}
\label{momentum}
\frac{\partial\rho\vec{v}}{\partial t}+\vec{\nabla}\cdot(\rho\vec{v}\vec{v})
   +\nabla p-\frac{(\vec{\nabla}\times\vec{B})\times\vec{B})}{4\pi}=+\rho\vec{g},
\end{equation}
\begin{equation}
\label{induction}
\frac{\partial\vec{B}}{\partial t}-\vec{\nabla}\times(\vec{v}\times\vec{B})=0,
\end{equation}
\begin{equation}
\label{energy}
\frac{\partial e}{\partial t}+\vec{\nabla}\cdot[(e+p)\vec{v}]=\rho\vec{g}\cdot\vec{v}-n^2\chi(T)-\nabla\cdot\vec{F_c},
\end{equation}
where $t$ is the time, $\rho$ the density,
$\vec{v}$ velocity, $p$ thermal pressure, $\vec{B}$ magnetic field,
$e$ the total energy,
$n$ number density, $\vec{F_c}$ the conductive flux according to \citet{Spitzer1962},
and $\chi(T)$ the radiative losses per unit emission measure \citep{Colgan2008}.
To close the set of Eqs. \ref{mass}-\ref{energy} we have a relation between internal, total, kinetic, and magnetic energy
\begin{equation}
\label{enercouple}
\frac{p}{\gamma-1}=e-\frac{1}{2}\rho\vec{v}^2-\frac{\vec{B}^2}{8\pi},
\end{equation}
where $\gamma=5/3$ denotes the ratio of specific heat and
the expression for solar gravitational acceleration
\begin{equation}
\label{solargravity}
\vec{g}=-\frac{G M_{\odot}}{r^2}\hat{r},
\end{equation}
where $G$ is the gravitational constant, $M_{\odot}$ denotes the mass of the Sun,
$r$ is the radial distance from the centre of the Sun,
and $\hat{r}$ is the corresponding unit vector.
In order to gain accuracy in the description of the thermal pressure,
we make use of the magnetic field splitting technique \citep{Powell1999},
as explained in detail in Sect. 2.3 of \citet{Pagano2013a}.
The radiative losses are treated using the exact integration method \citep{VanMarleKeppens2011}.

\begin{figure}[!htcb]
\centering
\includegraphics[scale=0.32,clip,viewport=120 40 900 630]{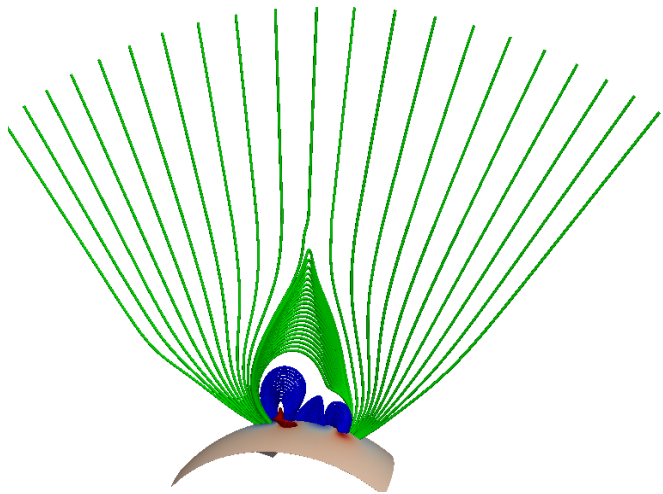}
\caption{Magnetic field configuration used as the initial condition in all the MHD simulations.
Red lines represent the flux rope, blue lines the arcades, and green lines the external magnetic field.
The lower boundary is coloured according to the polarity of the magnetic field
from blue (negative) to red (positive) in arbitrary units.}
\label{initialmagnetic}
\end{figure}

We start from the results of \citet{Pagano2013a} and \citet{Pagano2013b} to set up the MHD simulation.
In particular, we adopt the same magnetic configuration that leads
to flux rope ejections in these studies.
The configuration of the magnetic field is taken from
Day 19 in the simulation of \citet{MackayVanBallegooijen2006A}.
\citet{Pagano2013a} in Sect. 2.2 explain in detail how the magnetic field distribution is imported from the
GNLFFF model of \citet{MackayVanBallegooijen2006A} to our MHD simulations.
In the present paper the simulation domain extends over $3$ $R_{\odot}$ in the radial dimension starting from
$r=R_{\odot}$. The colatitude, $\theta$, spans from $\theta=30^{\circ}$ to
$\theta=100^{\circ}$ and the longitude, $\phi$, spans over $90^{\circ}$.
This domain extends to a larger radial distance than the domain used in \citet{MackayVanBallegooijen2006A}
from which we import the magnetic configuration.
To define the magnetic field for $r>2.5$ $R_{\odot}$, we assume it to be purely radial ($B_{\theta}=B_{\phi}=0$) where
the magnetic flux is conserved:
\begin{equation}
\label{brover25r}
B_r(r>2.5 R_{\odot},\theta,\phi)=B_r(2.5 R_{\odot},\theta,\phi)\frac{2.5^2}{r^2}.
\end{equation}
Figure \ref{initialmagnetic} shows a 3D plot of the initial magnetic configuration.
The flux rope (red lines) lies in the $\theta$ direction.
The flux rope is close to the point where an eruption will occur as it 
can no longer be held down by the overlying arcades.
The arcades are shown by the blue lines above which lies the
external magnetic field lines (green lines).
Some of the external magnetic field lines belong to the external arcade while some are open.

The boundary conditions of the MHD simulation are treated with a system of ghost cells.
Open boundary conditions are imposed at the outer boundary, reflective boundary conditions are set at the $\theta$ boundaries,
and the $\phi$ boundaries are periodic.
The $\theta$ boundary condition prevents any plasma or magnetic flux passing through,
while the $\phi$ boundary condition allows the plasma and magnetic field to freely evolve across the boundary.
These boundary conditions match those used in \citet{MackayVanBallegooijen2006A}.
In our simulations the flux rope only interacts with the $\theta$ and $\phi$ boundaries
near the end of the simulations, thus they do not affect our main results
regarding the initiation and propagation of the CME.
At the lower boundary we impose a fixed boundary condition taken from the first four $\theta$-$\phi$
planes of cells derived from the GNLFFF model.
The computational domain is composed of $256\times128\times128$ cells distributed in a uniform grid.

In \citet{Pagano2013a} the production of the initial condition
from the GNLFFF simulation and the dynamics of the ejection are discussed in detail,
while \citet{Pagano2013b} identified the parameter space where the flux rope is fully ejected.
In the present work the simulation parameters are fixed to generate a CME.
In particular, \citet{Pagano2013b} have shown that having a low-$\beta$ region above the flux rope
is a critical factor for the occurrence of the flux rope ejection.
As our aim is to synthesise realistic EUV observations, we set up a simulation
with realistic values for plasma density and temperature.
The maximum intensity of the magnetic field is set to $B_{max}=42$ $G$.
In order to do so, we prescribe a non isothermal solar corona.
To specify the temperature distribution we use the function $T(\vec{B})$,
\begin{equation}
\label{temperaturedistr}
T(\vec{B})=[(\frac{6}{2+(B_{\theta}/|B|)}-2)(T_{out}-T_{min})+T_{min}](1-G(\vec{B}))+T_{out}G(\vec{B}),
\end{equation}
where $T_{out}$ is the value of $T(\vec{B})$ when $B_{\theta}=0$ and $G(\vec{B})=0$. 
The parameter $T_{min}$ determines the minimum allowed value for $T(\vec{B})$ where $B_{\theta}=|B|$ and $G(\vec{B})=0$.
The function $G(\vec{B})$ is used to bound the dependence of $T$ on $\vec{B}$
to the regions where the magnetic field strength is significant.
The form applied in the present paper is
\begin{equation}
\label{gausstemperature}
G(\vec{B})=e^{-\frac{|B|^2}{2B_*^2}},
\end{equation}
where $B_*=3$ $G$ in our simulation.
While the choice of this ad hoc analytic formula may seem strange,
it is justified by the fact that in our set up the flux rope lies in the $\theta$ direction,
with a positive $B_{\theta}$.
It is the only structure with a strong shear in the initial condition of our simulation.
The form applied in Eq.\ref{temperaturedistr} allows us to produce
a cool dense region at the location of the flux rope (i.e. high $B_{\theta}$).
In principle, it is possible to generalise this temperature distribution
by replacing the $\theta$ direction with
the direction of the flux rope axis.
However, for the present simulations this would have little effect.

The thermal pressure distribution is independently specified using the solution for hydrostatic equilibrium
with uniform temperature set equal to $T_{out}$,
\begin{equation}
\label{pressurestratification}
p=\frac{\rho_{LB}}{\mu m_p}k_b 2 T_{out} e^{-{\frac{M_{\odot}G \mu m_p}{2 T_{out} k_b R_{\odot}}}} e^{{\frac{M_{\odot}G \mu m_p}{2 T_{out} k_b r}}},
\end{equation}
where $\rho_{LB}$ is the density at $r=R_{\odot}$ when $|B|=0$,
$\mu=1.31$ is the average particle mass in the solar corona,
$m_p$ is the proton mass, and
$k_b$ is Boltzmann constant.
Finally, the density is simply given by the equation of state applied to Eq.\ref{temperaturedistr} and Eq.\ref{pressurestratification}:
\begin{equation}
\label{eos}
\rho=\frac{p}{T(\vec{B})}\frac{\mu m_p}{k_b}.
\end{equation}

\begin{figure}[!htcb]
\centering
\includegraphics[scale=0.50]{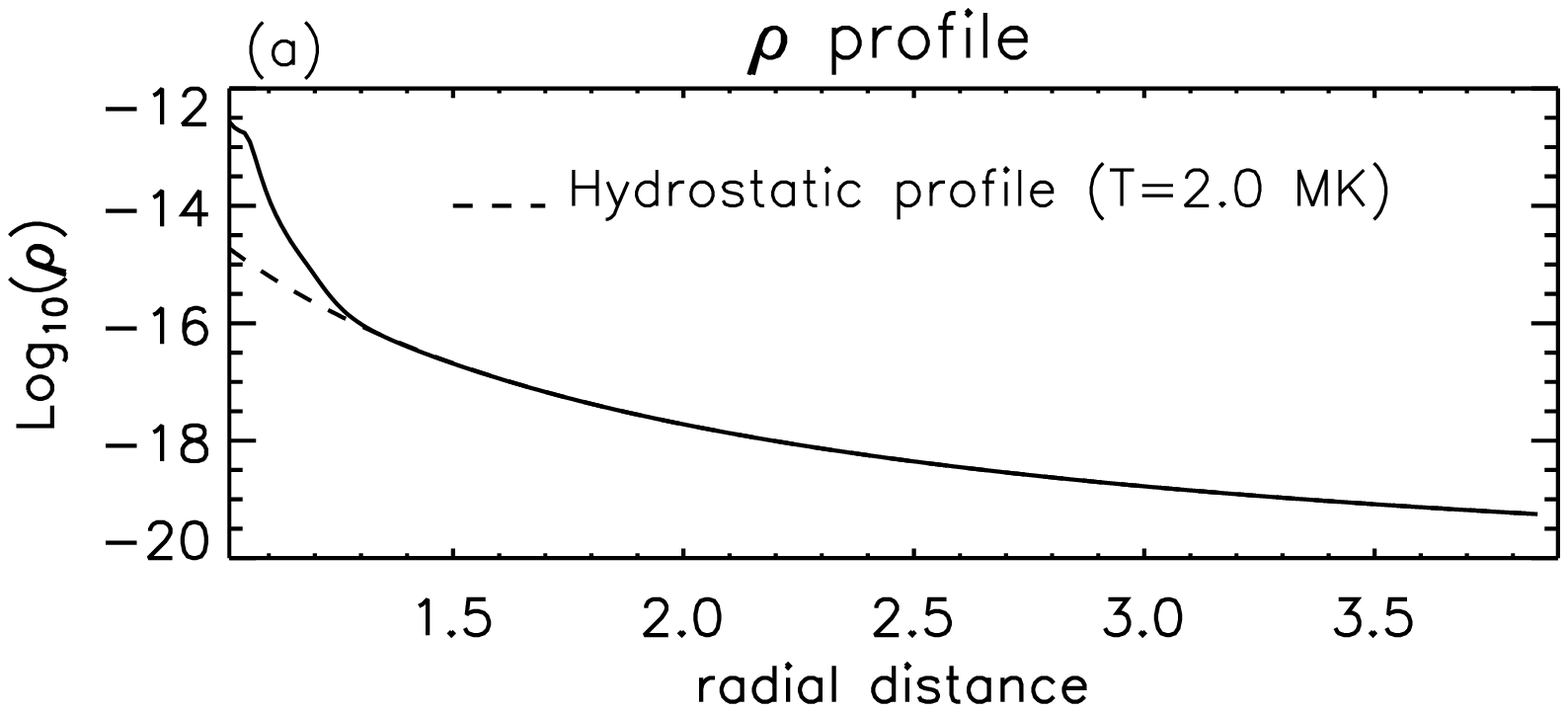}
\includegraphics[scale=0.50]{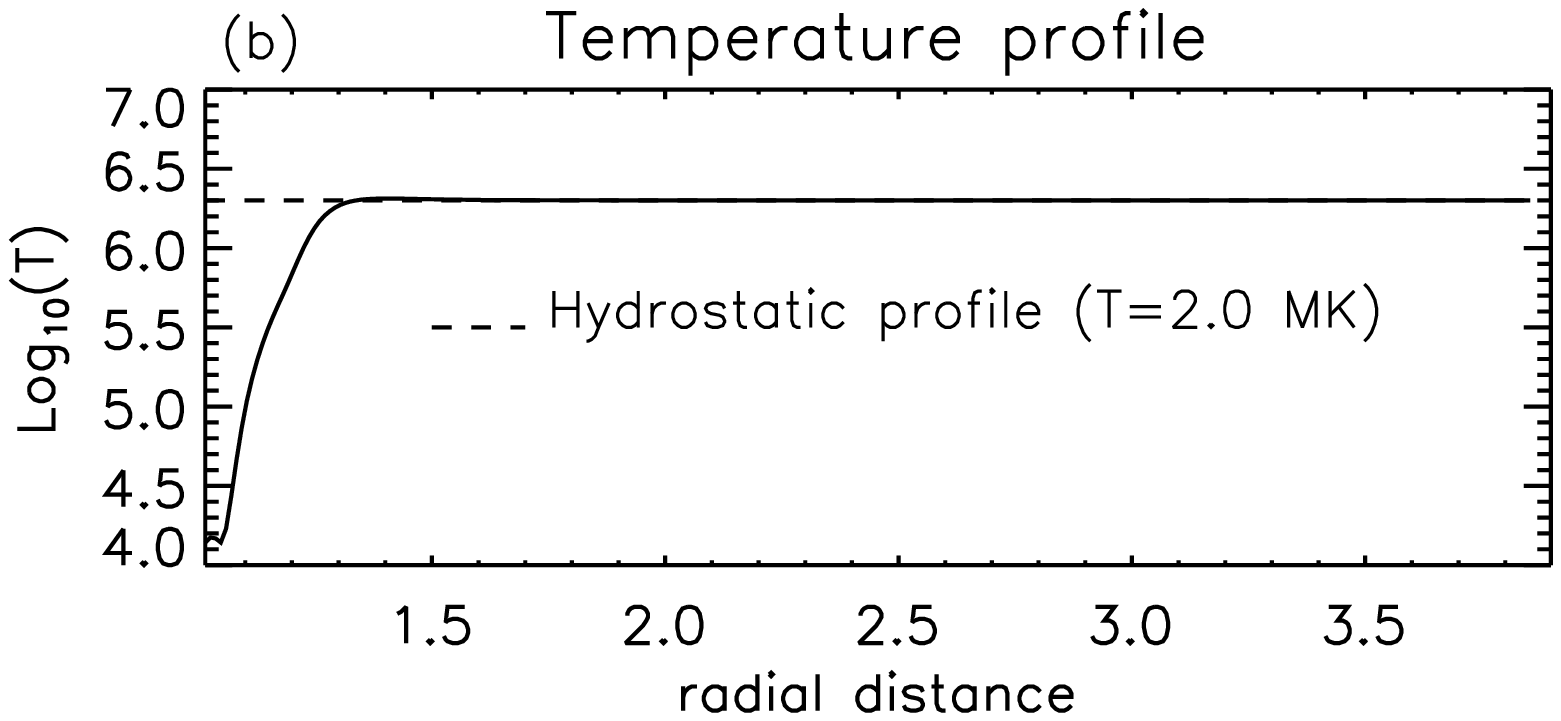}
\includegraphics[scale=0.50]{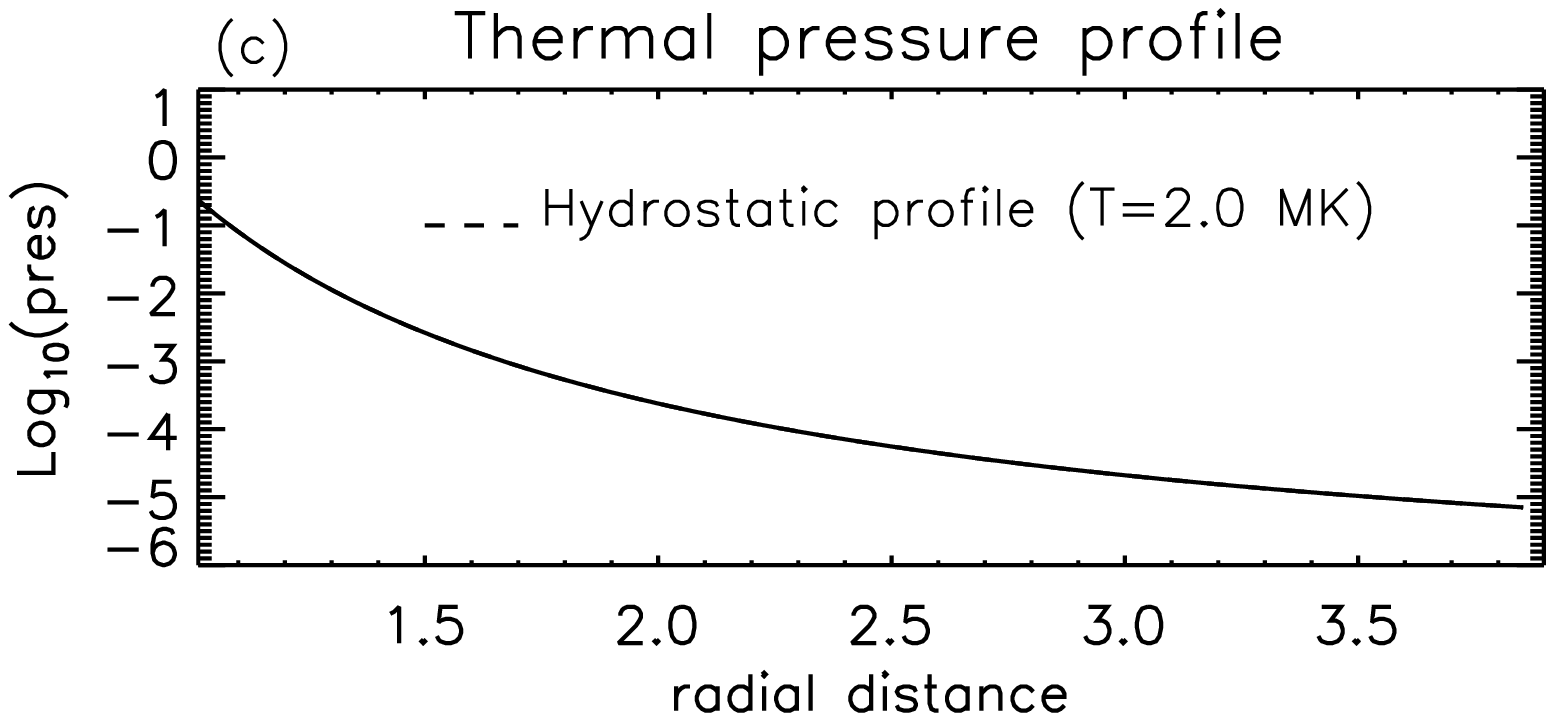}
\includegraphics[scale=0.50]{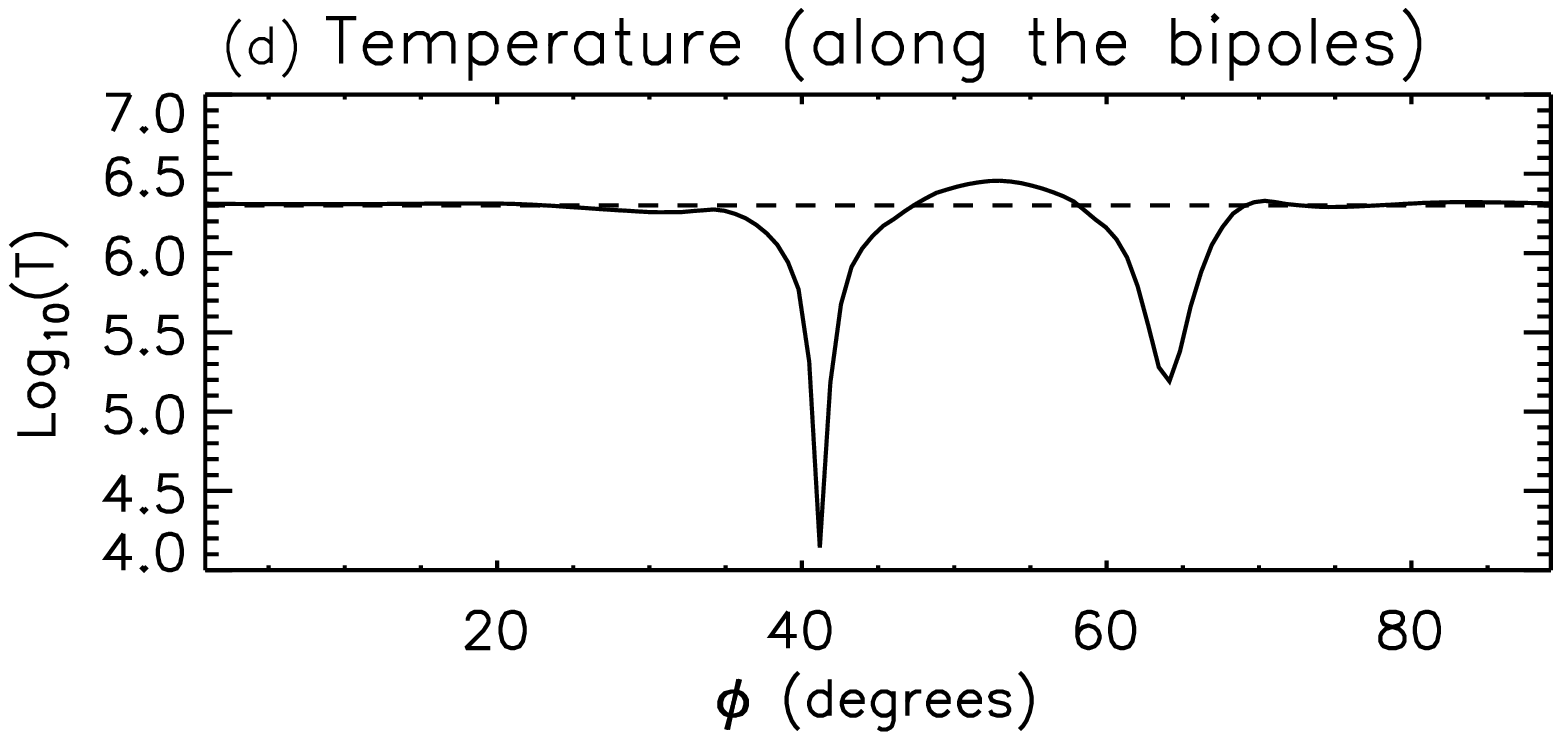}
\caption{Profiles along the radial direction above the centre of the LHS bipole
at $t=0$ $min$ of (a) $Log_{10}(\rho)$, (b) $Log_{10}(T)$, (c) $Log_{10}(p)$.
(d) Profile along the $\phi$ direction across the centre of the bipoles of $Log_{10}(T)$.
The dashed lines show the equivalent profiles for 
a coronal atmosphere in hydrostatic equilibrium at $T=2$ $MK$.}
\label{ptrhoprofile}
\end{figure}

In our simulation we choose
$T_{out}=2$ $MK$, 
$T_{min}=10^4$ $K$, and
$\rho_{LB}=3.5\times10^{-15}$ $g/cm^3$.
With these values, we obtain the atmospheric profile shown in Fig.\ref{ptrhoprofile}.
The first three plots show a radial cut of density, thermal pressure, and temperature (solid lines) from
the lower boundary to the external boundary passing through the centre of the left hand side (LHS) bipole
(where the flux rope lies).
In addition the temperature profile along a cut across the centre of the bipoles at the lower boundary is also shown.
For comparison, in each graph the corresponding hydrostatic solution at $T=2$ $MK$ is shown (dashed line).
The density profile (Fig.\ref{ptrhoprofile}a) shows a decreases due to gravitational stratification.
However, below $r\sim1.3$ $R_{\odot}$ there is an excess of density with respect to the hydrostatic profile.
This is due to the presence of the flux rope above the LHS bipole
which is prescribed denser and cooler than the surroundings in our model.
The temperature profile (Fig.\ref{ptrhoprofile}b) shows a varying temperature distribution.
The flux rope at $T\sim10^{4}$ $K$ is cooler than the surroundings,
where the external corona above the flux rope lies at $T=2$ $MK$.
While Fig.\ref{ptrhoprofile}b shows a radial cut above the flux rope,
the lower boundary of the simulation is mostly at $T=2$ $MK$
except for the locations around the polar inversion line of the bipoles where
the temperature is lower (Fig.\ref{ptrhoprofile}d).
The LHS bipole shows a much lower temperatature than the righ hand side (RHS) bipole,
because on the LHS bipole there is a fully formed flux rope,
which has a more prominent axial field component, $B_{\theta}$.
The thermal pressure profile (Fig.\ref{ptrhoprofile}c) is built to correspond to the hydrostatic profile.

It should be noted that Eqs. \ref{temperaturedistr}-\ref{eos}
do not produce a hydrostatic equilibrium
because of the varying temperature dependence on $B_{\theta}$.
In contrast, the thermal pressure profile assumes a constant temperature at $T=T_{out}$.
Such an atmosphere, if left to evolve,
would locally expand or contract depending on whether the temperature is larger or smaller than $T_{out}$.
However, as we have shown in Appendix A of \cite{Pagano2013b},
this has little or no effect on our simulation,
as its dynamics is entirely governed by the out-of-equilibrium magnetic field configuration
when the density profile departs so little from the pure gravitational profile.
This configuration immediately leads to the ejection of the flux rope
which results in a major perturbation of the whole coronal environment before
any significant motion due to gravity or thermal pressure gradients occur.
We note that the use of 
this atmosphere, which is not in hydrostatic equilibrium,
is necessary to produce the $\rho$ and $T$
distributions required for realistic emission images.

In conclusion, our atmosphere is suitable to realistically describe the region surrounding a flux rope.
While this is the case the values of density and thermal pressure appear too high
for an outer corona that is in equilibrium, partly because
we do not consider a solar wind that would significantly change
the outer coronal values of density and thermal pressure.
However, the plasma properties in the external corona are not important
for the present study which considers
the violent flux rope propagation and expansion in the low corona.

\subsection{Evolution of the MHD simulation}
\begin{figure}[!htcb]
\centering
\includegraphics[scale=0.21,clip,viewport=27 20 530 325]{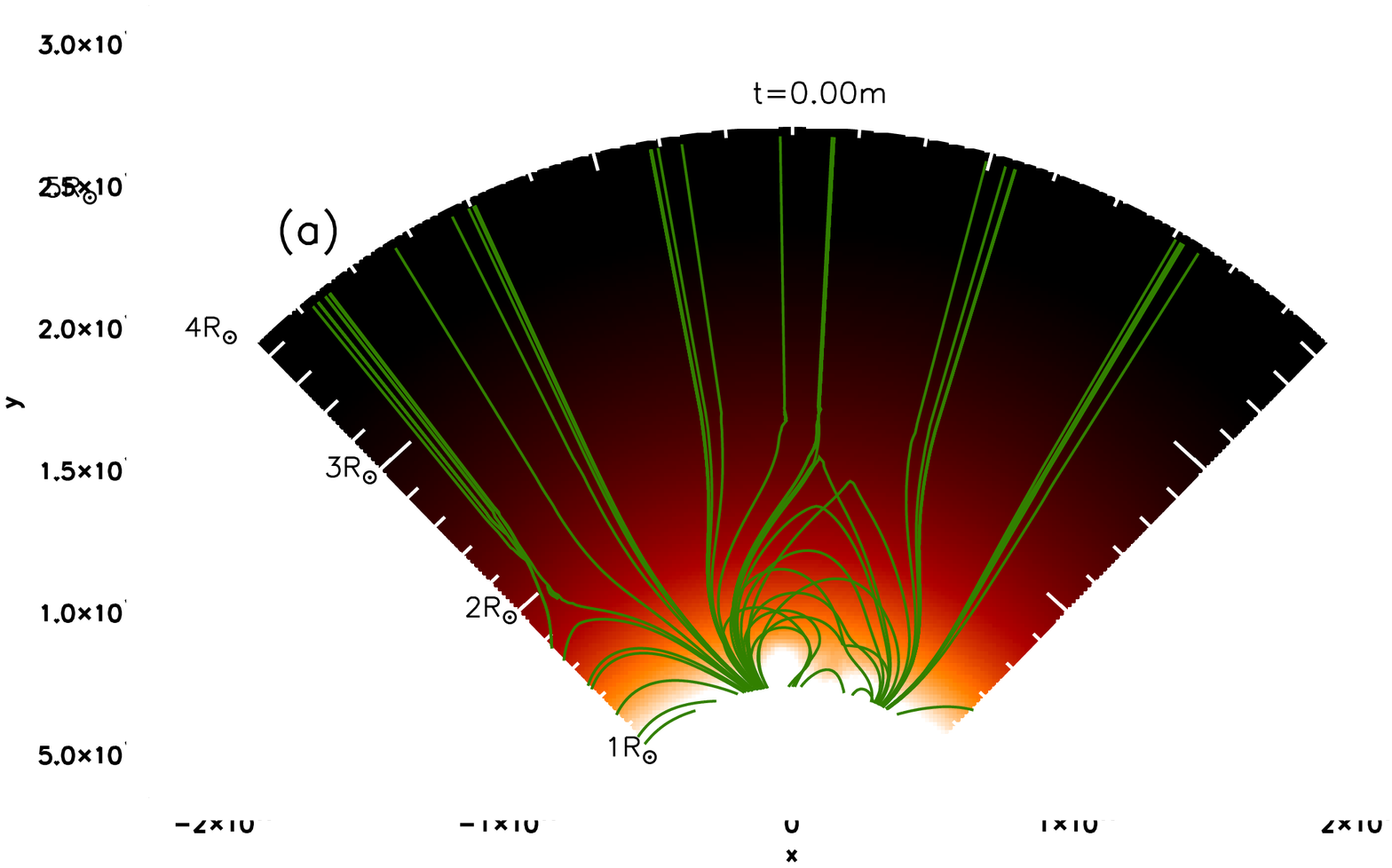}
\includegraphics[scale=0.18,clip,viewport=580 5 670 375]{fig3a.ps}
\includegraphics[scale=0.21,clip,viewport=27 20 530 325]{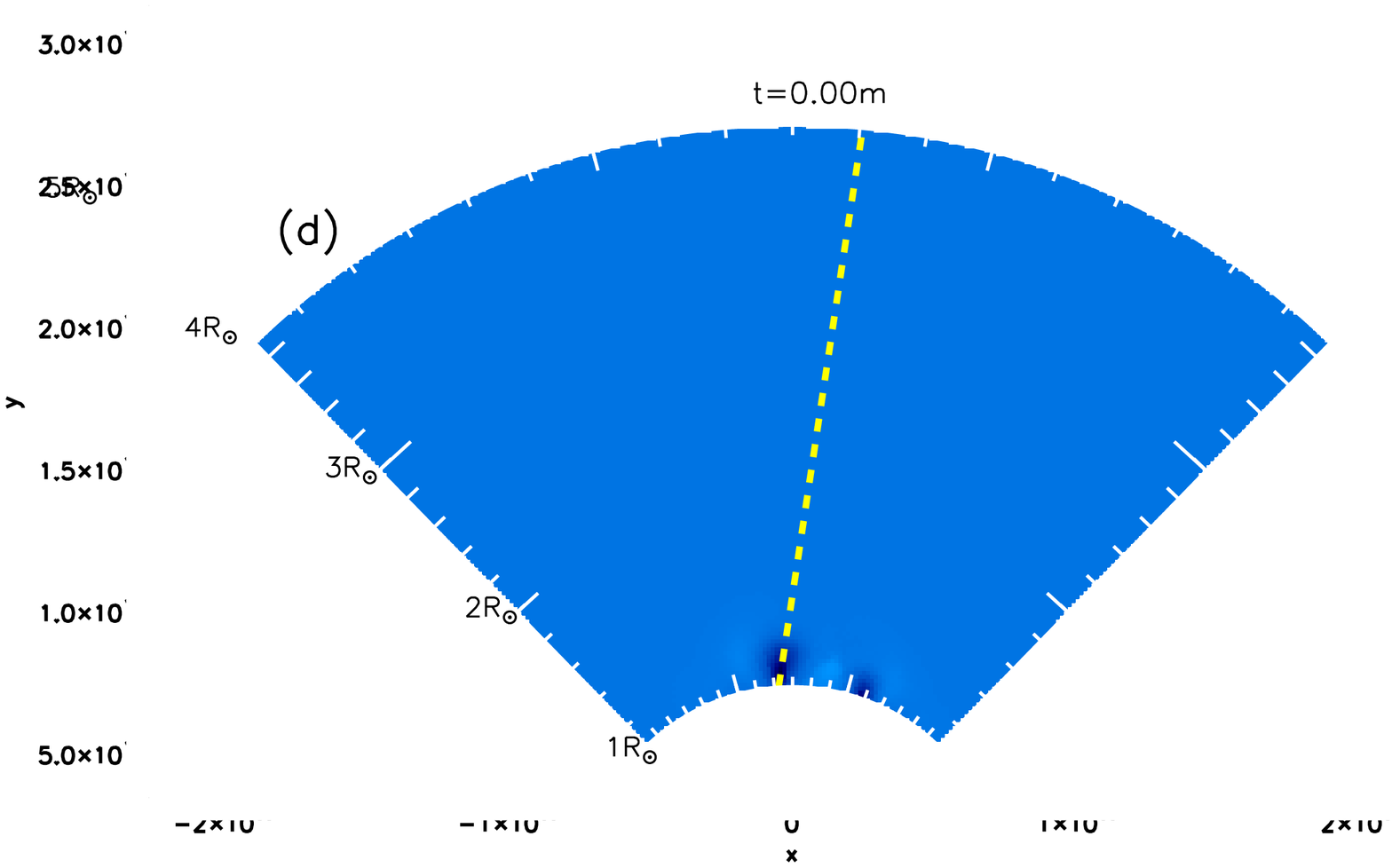}
\includegraphics[scale=0.18,clip,viewport=580 5 670 375]{fig3d.ps}

\includegraphics[scale=0.21,clip,viewport=27 20 530 325]{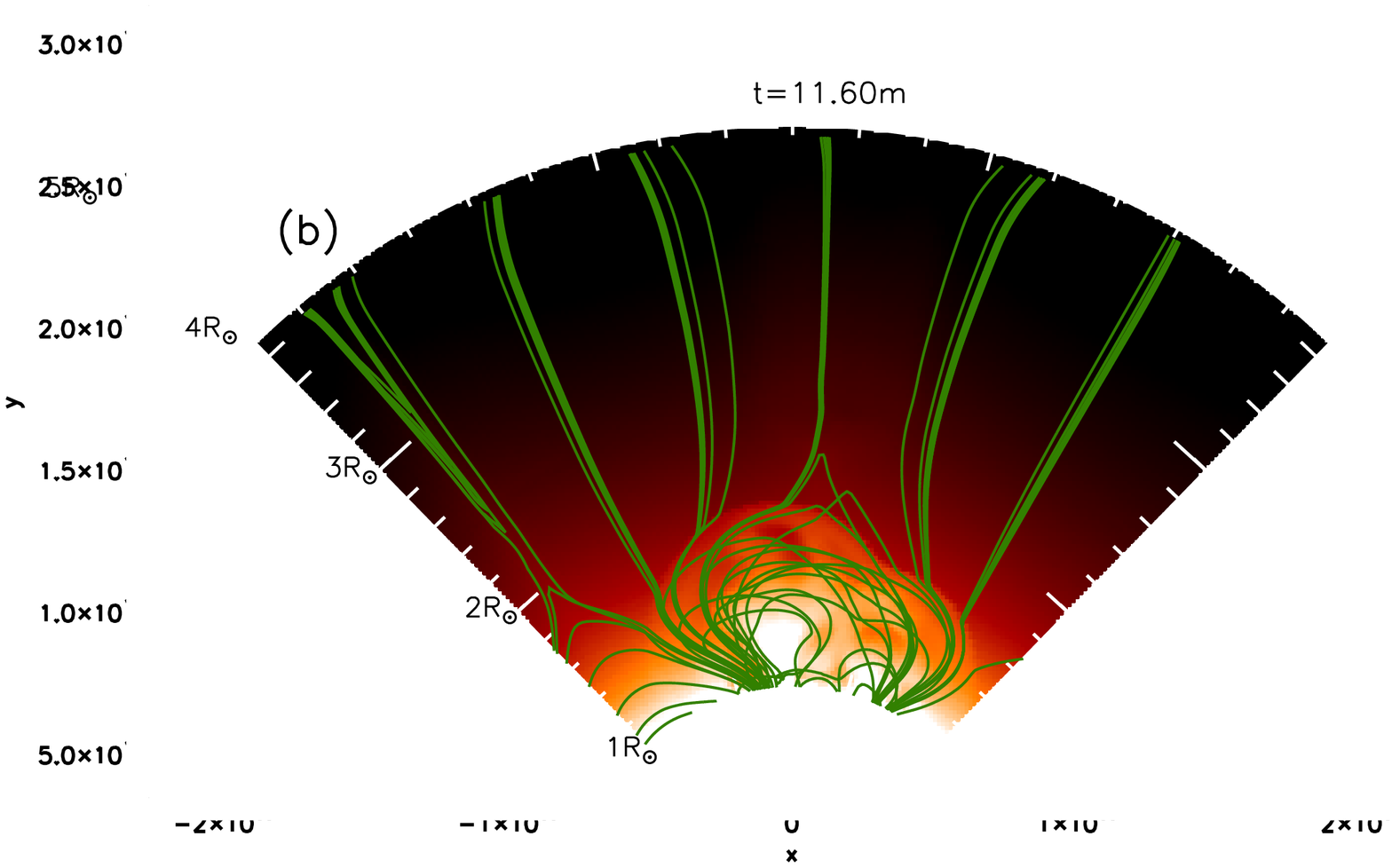}
\includegraphics[scale=0.18,clip,viewport=580 5 670 375]{fig3b.ps}
\includegraphics[scale=0.21,clip,viewport=27 20 530 325]{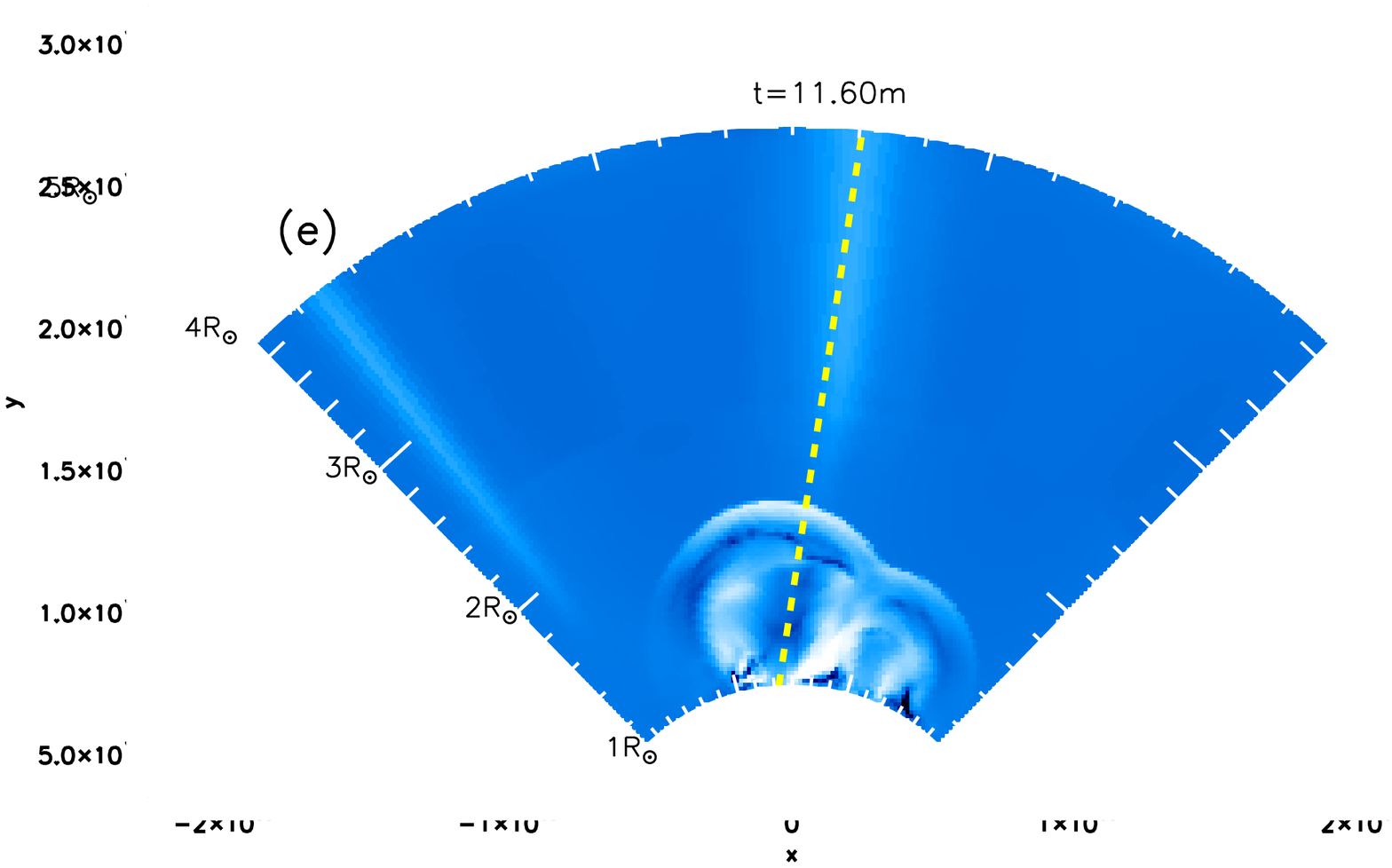}
\includegraphics[scale=0.18,clip,viewport=580 5 670 375]{fig3e.ps}

\includegraphics[scale=0.21,clip,viewport=27 20 530 325]{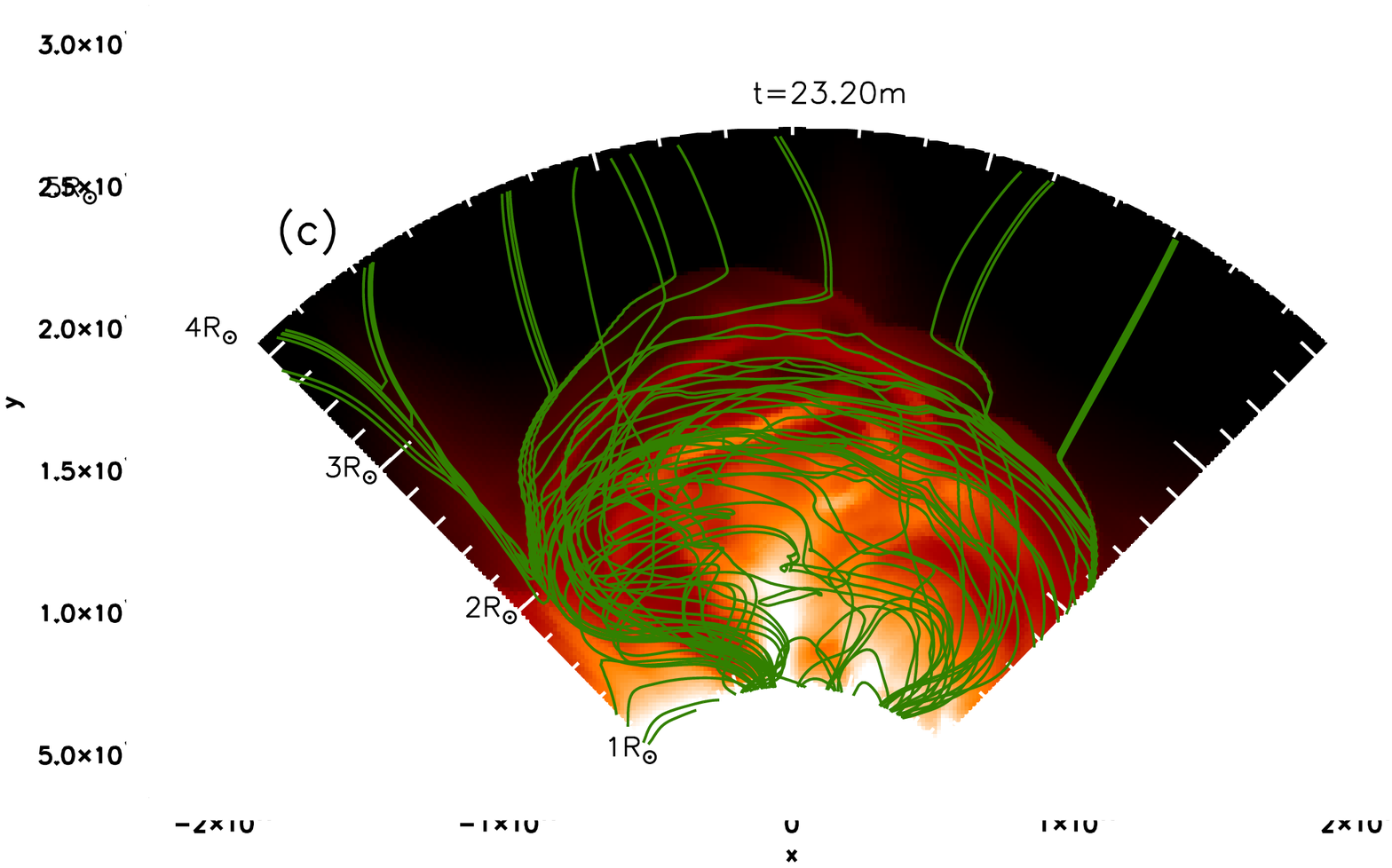}
\includegraphics[scale=0.18,clip,viewport=580 5 670 375]{fig3c.ps}
\includegraphics[scale=0.21,clip,viewport=27 20 530 325]{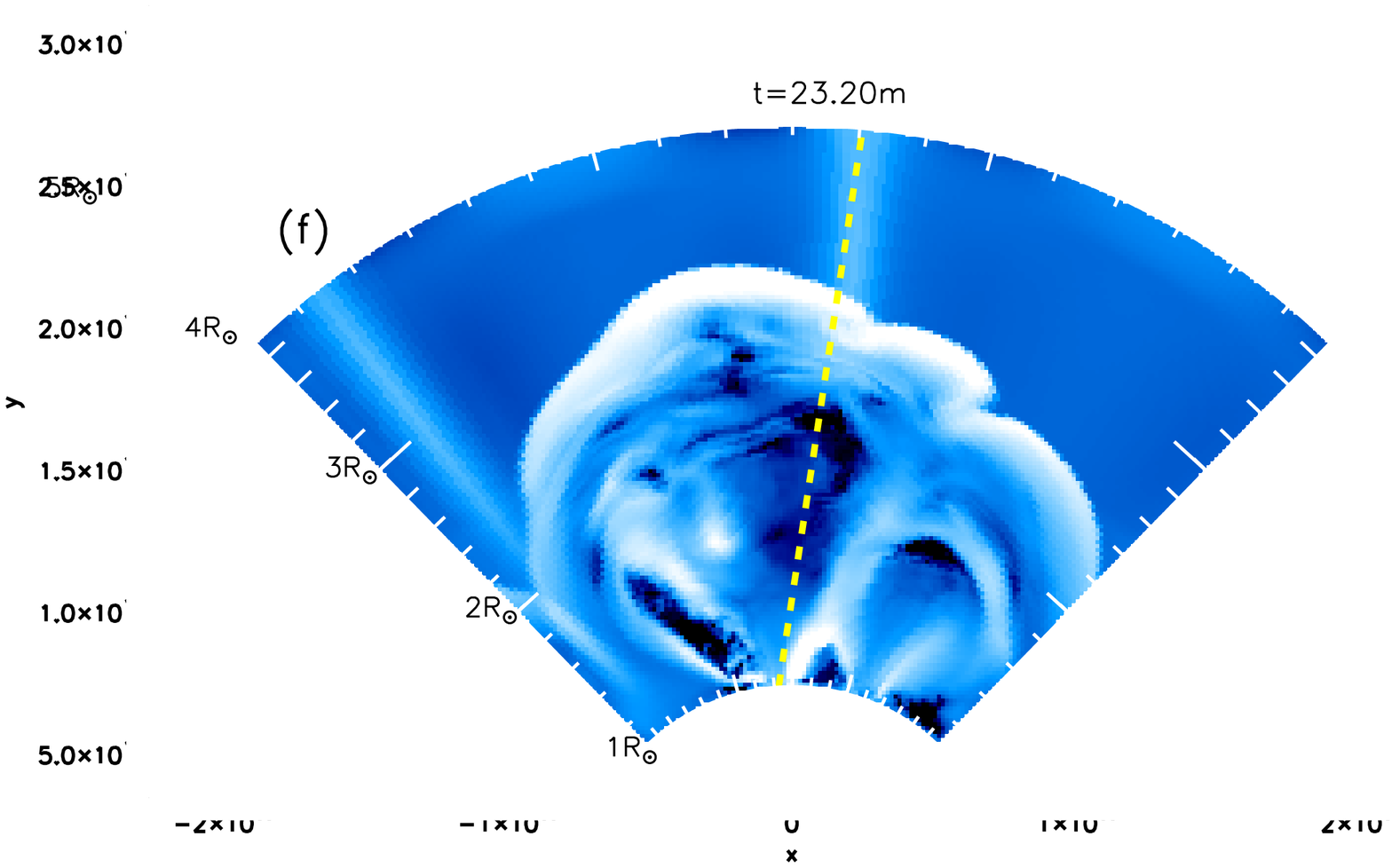}
\includegraphics[scale=0.18,clip,viewport=580 5 670 375]{fig3f.ps}
\caption{(a)-(c) Maps of $Log_{10}(\rho)$ in the $(r-\phi)$ plane
passing through the centre of the bipoles at $t=0$, $t=11.60$, and $t=23.20min$.
Superimposed are magnetic field lines plotted from the same starting points (green lines).
(d)-(f) Maps of $Log_{10}(T)$ on the same plane and at the same times.
Maps show the full domain of our simulations from $r=1$ $R_{\odot}$ to $r=4$ $R_{\odot}$.
The yellow dashed line is the cut for the plots in Fig.\ref{ptrhoprofile} and  Fig.\ref{cutevol}.
The temporal evolution is available in the online edition.}
\label{evolT2B6cdFR}
\end{figure}

The MHD simulation shows a very similar evolution to those already analysed in \citet{Pagano2013b}.
The evolution is illustrated in Fig.\ref{evolT2B6cdFR} where maps of density and temperature 
are shown in the ($r-\phi$) plane passing through the centre of the bipoles
(a movie of Fig.\ref{evolT2B6cdFR} is available in the online edition).
Initially the flux rope lies near the lower boundary
and as soon as the system is allowed to evolve it is ejected radially outwards.
Once the ejection occurs the high density region rises.
Initially the high density region remains in the shape of the ejected flux rope (Fig.\ref{evolT2B6cdFR}b);
however, near the end of the simulation it is less identifiable as a flux rope (Fig.\ref{evolT2B6cdFR}c).
After approximately 30 minutes the front of the high density front reaches the outer boundary at $4$ $R_{\odot}$.

The magnetic field configuration undergoes a major evolution and reconfiguration as a result of the ejection.
While this occurs, the flux rope can still be identified during the entire simulation
by the presence of a strong axial component of the magnetic field
($B_{\theta}$ - not shown; see \citet{Pagano2013b}, Sect. 3.1).
The magnetic flux is expelled outwards and at $t=23.20min$ (Fig.\ref{evolT2B6cdFR}c)
a region of compressed magnetic field is visible ahead of the ejection,
where the front of the ejection compresses both plasma and magnetic flux.

The flux rope is initially colder than the surrounding corona at approximately $T=10^4$ $K$
(Fig.\ref{evolT2B6cdFR}d, dark area).
At the initial stage of the ejection
its temperature increases and
a region temperature around $T=1$ $MK$ is produced.
This region then expands and propagates upwards
and the density starts decreasing along with the temperature.
At the same time a hot front heated by compression propagates ahead of the flux rope,
where the temperature is about $10^7$ $K$.
From Fig.\ref{evolT2B6cdFR}e it can be seen that
the temperature distribution at the location of the ejection
presents some variations, but a clear hot front overlies the ejection.
This front clearly marks where the perturbation of the ejection has reached (Fig.\ref{evolT2B6cdFR}f).

As explained in \citet{Pagano2013a}, the flux rope is initially ejected radially outwards
because of an excess of the radial Lorentz force underneath.
However, immediately afterwards, the ejection propagates non-radially
in the direction of the null point which lies above the arcade system
(yellow dashed line in Figs. \ref{evolT2B6cdFR}d-f),
because the confining Lorentz force exterted by the magnetic arcades is weakest there 
and so this becomes a favourable escape direction for the flux rope.
This is particularly visible in Figs. \ref{evolT2B6cdFR}d-f,
where the colder plasma moves along the yellow dashed line.

\begin{figure}[!htcb]
\centering
\includegraphics[scale=0.37]{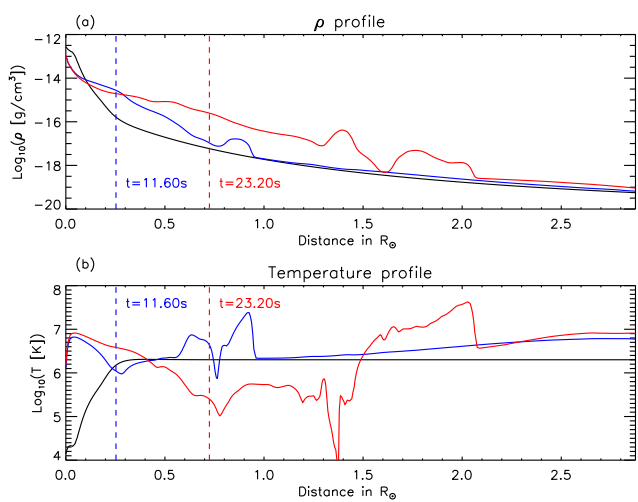}
\caption{(a) Profiles of $Log_{10}(\rho)$.
(b) Profiles of $Log_{10}(T)$.
All the profiles are along the yellow dashed line in Fig.\ref{evolT2B6cdFR}d
which represents the direction of the flux rope ejection
at $t=0$ $min$ (black line), $t=11.60$ $min$ (blue line), and $t=23.20$ $min$ (red line).
The dashed lines mark the position of the centre of the flux rope at each time.}
\label{cutevol}
\end{figure}

In order to accurately follow the density and temperature evolution,
we show in Fig.\ref{cutevol} the profile of density and temperature above the centre of the 
LHS bipole along the direction of ejection (yellow cut in Fig.\ref{evolT2B6cdFR}d) at $t=0$, $11.60$, and $23.20$ $min$.
At each time, the position of the centre of the flux rope is shown with a dashed vertical line.
The centre of the flux rope is located where the quantity $B_{\theta}/|B|$
is highest along the considered cut.
In our framework, this identifies where the strongest axial component of the flux rope lies.

At all times, the flux rope lies where an excess in density is present (Fig.\ref{cutevol}a).
The radial extension of the excess density region is smallest at $t=0min$,
and increases with time.
At $t=11.60min$ the flux rope extends over $0.5$ $R_{\odot}$.
Simultaneously, the flux rope density decreases from $\rho\sim10^{-12.5}$ $g/cm^3$
to $\rho\sim10^{-15}$ $g/cm^3$.
A dense front is located ahead of the propagating flux rope at larger radial distance.
This front does not exist at $t=0min$ as it is a consequence of the flux rope propagation that compresses plasma ahead of it.
Throughout the simulation the density of the front is around $\rho\sim10^{-17}$ $g/cm^3$.

\begin{figure}[!htcb]
\centering
\includegraphics[scale=0.37]{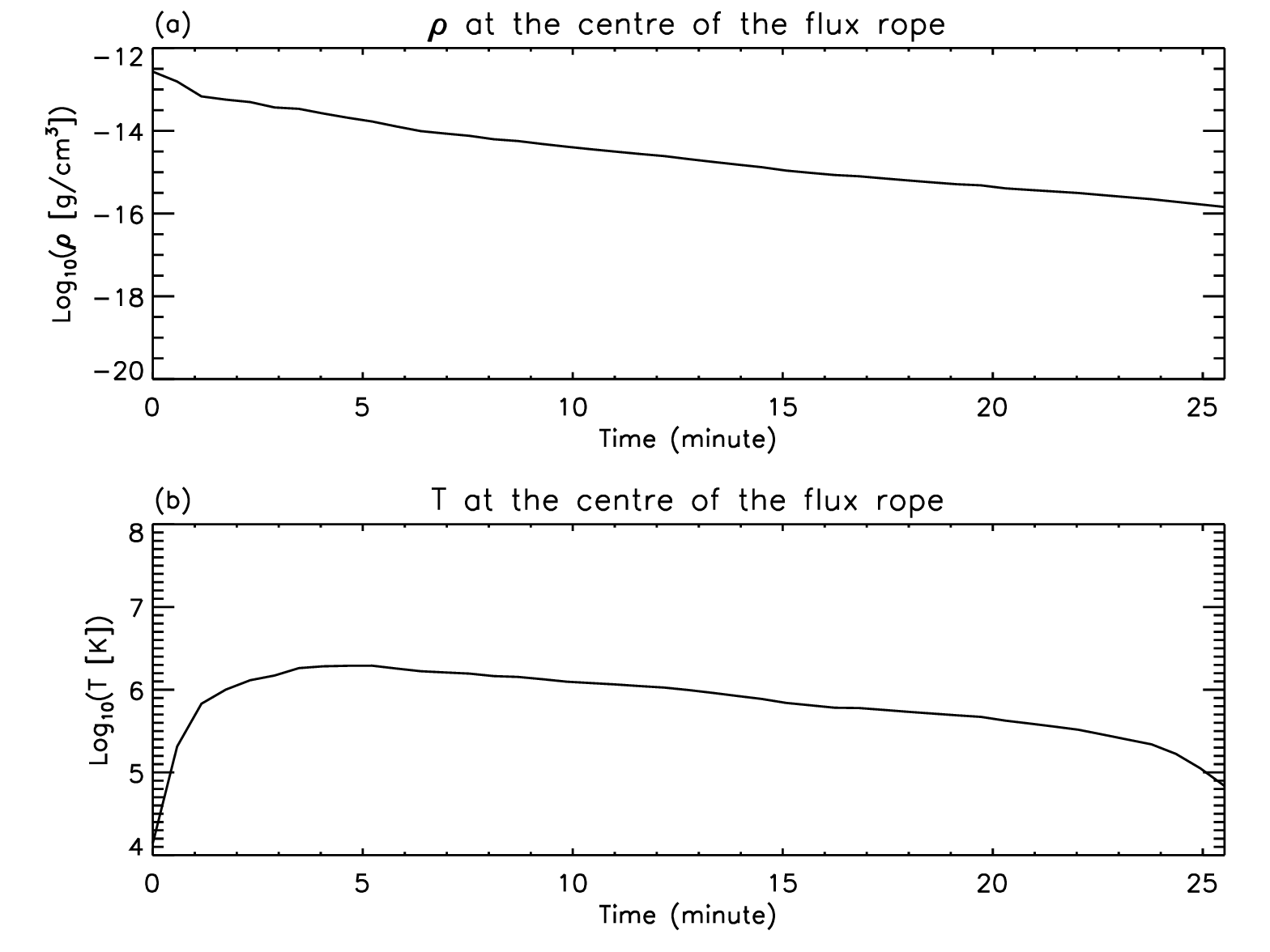}
\includegraphics[scale=0.37]{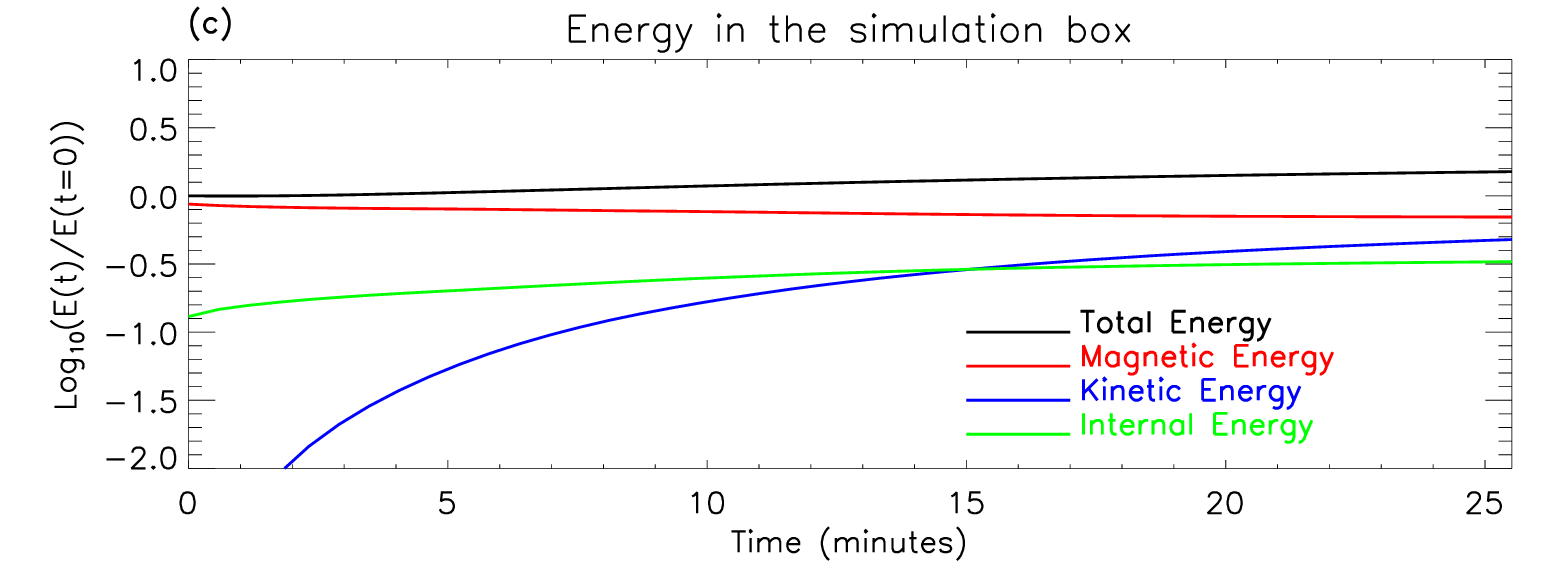}
\caption{(a) Value of $Log_{10}(\rho)$ at the centre of the flux rope as a function of time.
(b) Value of $Log_{10}(T)$ at the centre of the flux rope as a function of time.
(c) Variation of each energy term in different colours integrated in the whole domain as function of time in units of the initial total energy.}
\label{cfrplots}
\end{figure}

Figure \ref{cutevol}b shows the evolution of temperature along the same cut,
where the flux rope is initially at $T=10^4$ $K$ and is colder than the surroundings.
By $t=11.60$ $min$ the temperature of the flux rope has increased to be around $T\sim10^6$ $K$
because of the conversion of magnetic energy into internal energy
as a result of magnetic reconnection due to numerical diffusion.
While the flux rope plasma is initially heated,
its temperature decreases to $T\sim10^{5.5}$ $K$ by $t=23.20$ $min$.
It should be noted that at all times the position of
the centre of the flux rope roughly corresponds to a dip in temperature.
The high density front seen in Fig.\ref{cutevol}a presents a very sharp gradient in temperature,
as the compression occurring there heats the plasma to beyond $T\sim10^7$ $K$.
In Fig.\ref{evolT2B6cdFR}e the linear white zone that extends up
above the main ejection along the yellow dashed line is heated by the conversion of magnetic energy into thermal energy,
because it lies along a separatrix surface.

Using the density and temperature at the centre of the flux rope
we may follow the thermodynamic evolution of the flux rope.
Figure \ref{cfrplots}a shows the evolution of the density at the centre of the flux rope as a function of time.
The density steadily decreases from $\rho\sim10^{-12}$ $g/cm^3$ to $\rho\sim10^{-16}$ $g/cm^3$ because of the expansion during the eruption.
The temperature (Fig.\ref{cfrplots}b) initially increases rapidly by two orders of magnitude in less than 5 minutes
and then drops again to less than $T\sim10^{5}$ $K$ as the expansion occurs.

Figure \ref{cfrplots}c shows the evolution of the total energy in the entire computational box over the simulation.
There is an increase in the total enery during the simulation
because the simulation does not take place in an isolated box.
The lower boundary conditions are fixed and as a result of the magnetic field in the ghost cells not being in equilibrium
a Poynting flux (thus energy) is injected into the simulation.
In addition, mass, energy, and magnetic flux are allowed to cross the upper boundary which leads to losses.
Such a phenomenon, although present, was not significant in our previous work \citep{Pagano2013a}
becasue of the different $\beta$ regime where the uniform pressure distribution led to
most of the energy being stored as internal energy.
While the energy does increase in the present simulation very similar results are found
in cases where it did not increase.
Therefore we do not believe that this significantly affects the results.

Throughout the simulation
most of the total energy is stored in magnetic energy.
Later on in the simulation, a portion of the magnetic energy 
is converted into internal and kinetic energy and thus the magnetic energy decreases.
The internal energy increases, to about one third of the magnetic energy,
but the simulation always develops in a mostly low-$\beta$ regime.
The kinetic energy is initially zero, but it quickly increases
to be greater than the internal energy and becomes stable at about 35\% of the total energy.
The code solves the MHD energy equation (Eq.\ref{energy}) by computing the partial time derivative of the total energy, $e$.
When magnetic energy is lost from one time step to another because of numerical resistivity,
or the same happens to kinetic energy because of numerical viscosity
this energy is automatically 
converted into thermal energy through Eq.\ref{enercouple}.

\begin{figure}[!htcb]
\centering
\includegraphics[scale=0.37]{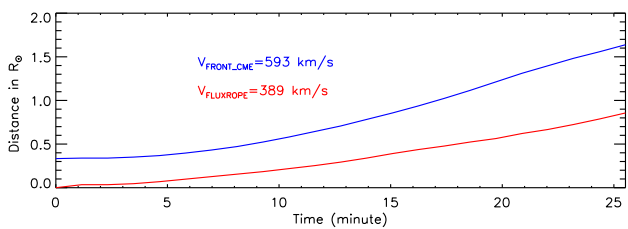}
\caption{Distance travelled by the centre and the top of the flux rope as a function of time.}
\label{centerfrposition}
\end{figure}

Finally, Fig.\ref{centerfrposition} shows the position of the centre of the flux rope
and the top of the flux rope as a function of time.
The parameters we have chosen for the simulation set conditions in which 
the flux rope is expelled from the low solar corona.
The flux rope undergoes an acceleration phase for the first 20 minutes
and then it travels outwards at approximatelty constant speed.
Whereas the average speed for the centre of the flux rope is $v\sim389$ $km/s$,
the top of the flux rope, corresponding to the CME front travels faster at $v\sim593$ $km/s$
because of the combined motion of propagation and expansion.

\subsection{Role of thermal conduction and radiative losses}

The simulation presented here aims to reproduce a realistic temperature profile in order to synthesise
the emission visible from AIA/SDO.
In order to do so, unlike our previous work, we have
used the versatility offered by MPI-AMRVAC to include non-ideal MHD terms such as
anisotropic thermal conduction and optically thin radiative losses in Eq.\ref{energy},
as successfully used and described in \citet{Fang2013} and \citet{Xia2012}.
We now consider the conquences of these terms.
Thermal conduction in the solar corona is extremely efficient along magnetic field lines and
since the thermalisation times are much shorter
than quasi-stable evolution times, it plays no role during equilibrium conditions.
Thus, the plasma is isothermal along a magnetic field line.
However, thermal conduction becomes very important for events such as CMEs where plasma is rapidly heated.
In our simulation, thermal conduction does not significantly affect the flux rope temperature,
as the flux rope is embedded in magnetic field lines where the thermal gradient is mostly perpendicular
to the magnetic field lines.
Thus, any heat exchange with the surroundings is prevented.
In contrast, thermal conduction does become effective at the hot front
where thermal gradients are parallel to magnetic field lines and 
the configuration is such as to allow heat flow.
The thermalisation timescale in the solar corona, $\tau_c$, in CGS unites is given by \citep{Pagano2007}
\begin{equation}
\tau_c=\frac{21 n k_b L^2}{18.4\times10^{-7} T^{2.5}},
\end{equation}
where L is a characteristic length scale
and $n$ is the number density.
Average estimations in our simulations are $L\sim R_{\odot}/2$, $n\sim10^7$ $cm^{-3}$, and $T\sim10^{6.4}$ $K$,
thus $\tau_c\sim1000$ $s$.
The presence of thermal conduction allows heat flow along the magnetic field lines
leading to a more uniform temperature profile along these magnetic field lines.
This occurs over timescales comparable to the dynamic timescales of our MHD simulation.
In contrast, radiative losses do not have a significant effect on either the hot front, or the flux rope.
The timescale of radiative losses is
\begin{equation}
\tau_r=\frac{3}{2}\frac{k_b T}{n P(T)},
\end{equation}
where $P(T)$ is the radiative losses per unit emission measure function \citep{Raymond1977}
and can be computed as $P(T)=10^{-17.73}T^{-2/3}$ in our temparature range \citep{Rosner1978}.
Thus, using flux rope values for the particle density $n$,
we have $\tau_r\sim10^4$ $s$.

\begin{figure}[!htcb]
\centering
\includegraphics[scale=0.21,clip,viewport=27 20 530 325]{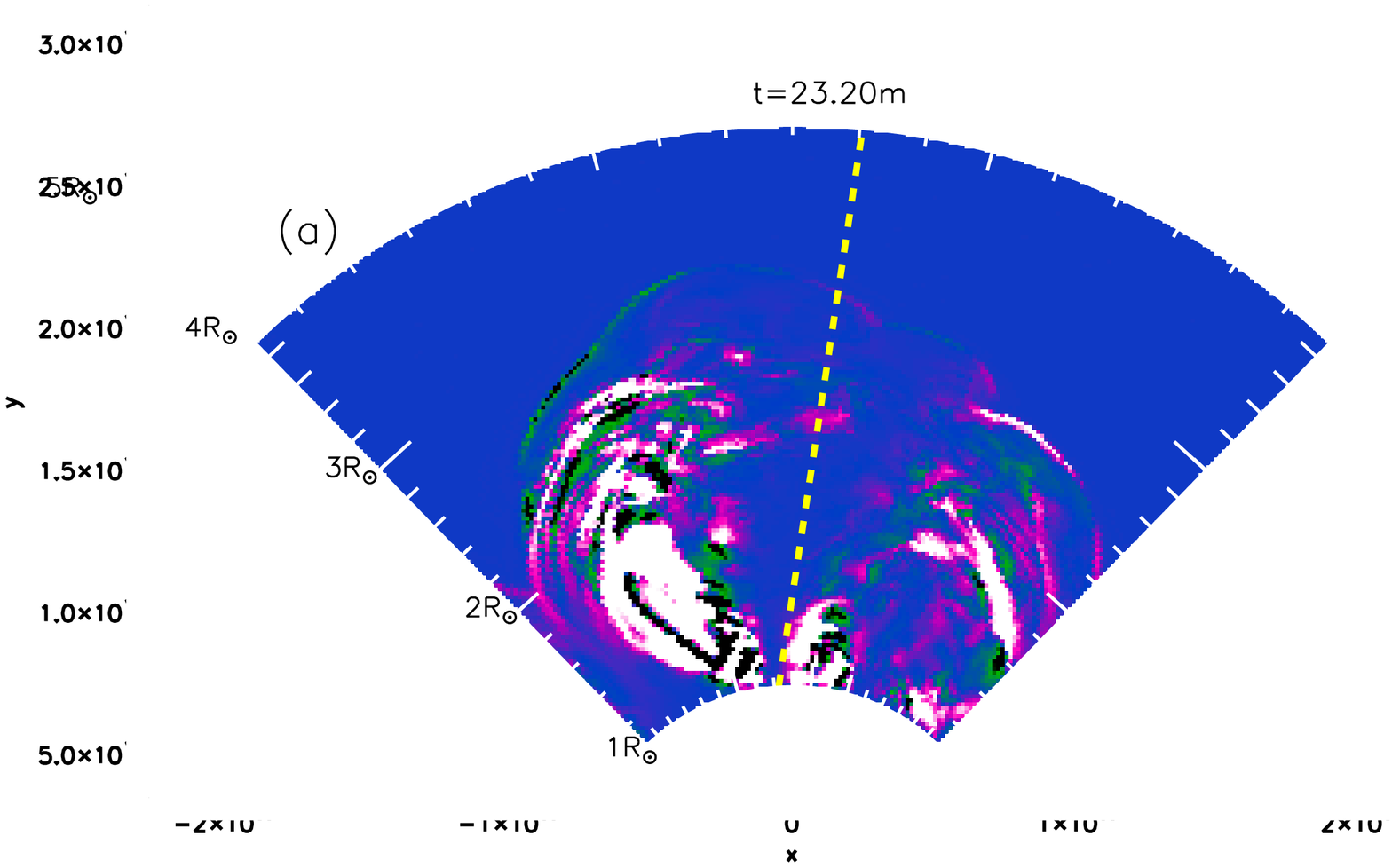}
\includegraphics[scale=0.18,clip,viewport=580 5 670 375]{fig7a.ps}
\includegraphics[scale=0.21,clip,viewport=27 20 530 325]{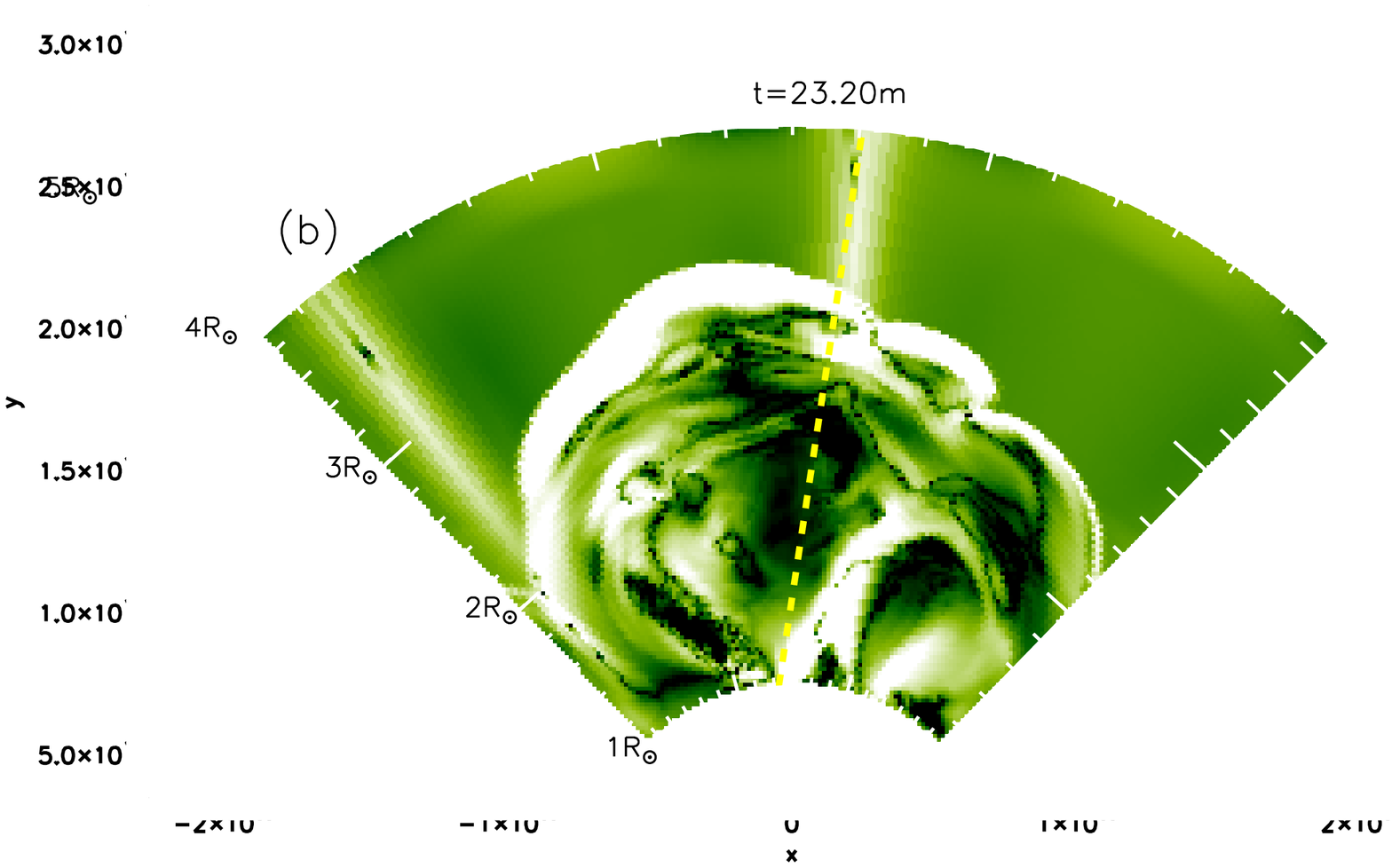}
\includegraphics[scale=0.18,clip,viewport=580 5 670 375]{fig7b.ps}
\caption{(a) Map of temperature differences between
the simulation where the non-ideal terms are considered and
the simulation without them in the $(r-\phi)$ plane
passing through the centre of the bipoles at $t=23.20min$.
(b) Map of $Log_{10}$ of $|\nabla T_{\parallel B}|$, the gradient along the magnetic field
on the same plane and at the same time.}
\label{tempdiff}
\end{figure}

In order to test these considerations we ran another simulation identical to the one presented in Sect. \ref{mhdsimulation},
but without the non-ideal MHD source terms in Eq.\ref{energy}.
Figure \ref{tempdiff}a shows the temperature difference between
the simulations with and without the non-ideal source terms.
The temperatures between the two models are significantly different
where the difference is typically about $1$ $MK$ (with localised peaks reaching $\sim20$ $MK$), higher or lower.
The difference in temperature shows that modelling
without thermal conduction and radiative losses
can both overestimate and underestimate the temperatures
by as much as $1$ $MK$.
It should be noted that significant differences are only evident
at the sides of the bipoles ($10^6$ $K$) and to a minor extent ahead of the flux rope ($10^5$ $K$).
The simulations with or without the non-ideal effects
do not show a significant variation at the flux rope location.
In the region of the flux rope motion
(along the yellow dashed cut) there is a good agreement 
between the two simulations within a temperature of about $5\times10^4$ $K$,
because the highly twisted magnetic field of the flux rope prevents heat exchange
and thermal condution is inefficienct there.
In Fig.\ref{tempdiff}b, we show the magnitude of the temperature gradient along the magnetic field lines
\begin{equation}
\label{nablatb}
|\nabla T_{\parallel B}|=|\frac{\nabla T\cdot\vec{B}}{|\vec{B}|}|
\end{equation}
for the simulation where non-ideal terms are considered.
It can be seen that the region where the gradient is lowest corresponds to the centre of the flux rope.
In contrast, the regions that show a greater temperature difference in Fig.\ref{tempdiff}a
mostly correspond to regions where the quantity $|\nabla T_{\parallel B}|$ is higher in Fig.\ref{tempdiff}b.

\section{AIA emission synthesis}
\label{aiasdoemissionsythesis}

We develop a simple technique that synthesises
the observations of AIA
from our MHD simulation with the use of the AIA module in Solar Software.
The synthesis is carried out as follows in three steps:
(i) with the AIA module of the Solar Software we create the instrumental response function for each channel as a function of temperature,
(ii) we then compute the synthesised emission from each cell of the MHD simulation (interpolated in a 3D cartesian domain), and
(iii) finally we integrate along the line of sight.
Therefore, we compute the emission from each element of plasma observed by each EUV channel of AIA as
\begin{equation}
EM_{ch.}(n,T)=n^2 \zeta(T)_{ch.},
\label{emch}
\end{equation}
where $\zeta(T)_{ch.}$ is provided by the AIA module in Solar Software.
\begin{figure}[!htcb]
\centering
\includegraphics[scale=0.50]{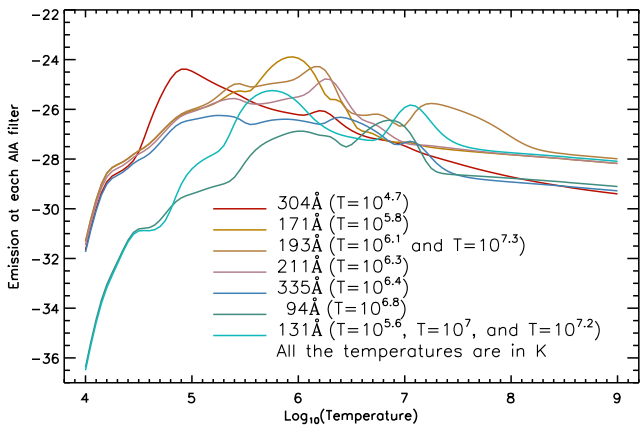}
\caption{Values of $Log_{10}(\zeta(n,T)_{ch.})$ for all the AIA channels as a function of $T$}
\label{allfilters}
\end{figure}
Figure \ref{allfilters} shows the functions $\zeta(T)_{ch.}$ for all the channels of AIA.
For all the channels the emission peak
lies far from our temperature extrema at $T=10^{4}$ $K$ and $T=10^{8}$ $K$.
The temperature range of the channels includes the temperature values
found in our simulation.

\section{Results}
\label{results}

In the present paper, we perform a MHD simulation of the ejection of a magnetic flux rope
and now we apply a simple emission model to synthesise
the corresponding observations of AIA.
In particular, we show here the synthesised observations for four of the AIA channels
($304\AA$, $171\AA$, $335\AA$, and $94\AA$).
These channels show the evolution of the flux rope ejection in the solar corona at commonly observed temperatures.
In Sect. \ref{coldchannels} we describe the synthesised observations for the channels at
$304\AA$ (peak at $T\sim10^{4.7}$ $K$) and $171\AA$ (peak at $T\sim10^{5.8}$ $K$), observing relatively cool plasma.
In Sect. \ref{hotchannels} we describe the synthesised observations for the channels at
$335\AA$ (peak at $T\sim10^{6.4}$ $K$) and $171\AA$ (peak at $T\sim10^{6.8}$ $K$), observing hot plasma.

\subsection{Cold channels: $304\AA$ and $171\AA$}
\label{coldchannels}

\begin{figure}[!htcb]
\centering
\includegraphics[scale=0.49,clip,viewport=045 385 315 665]{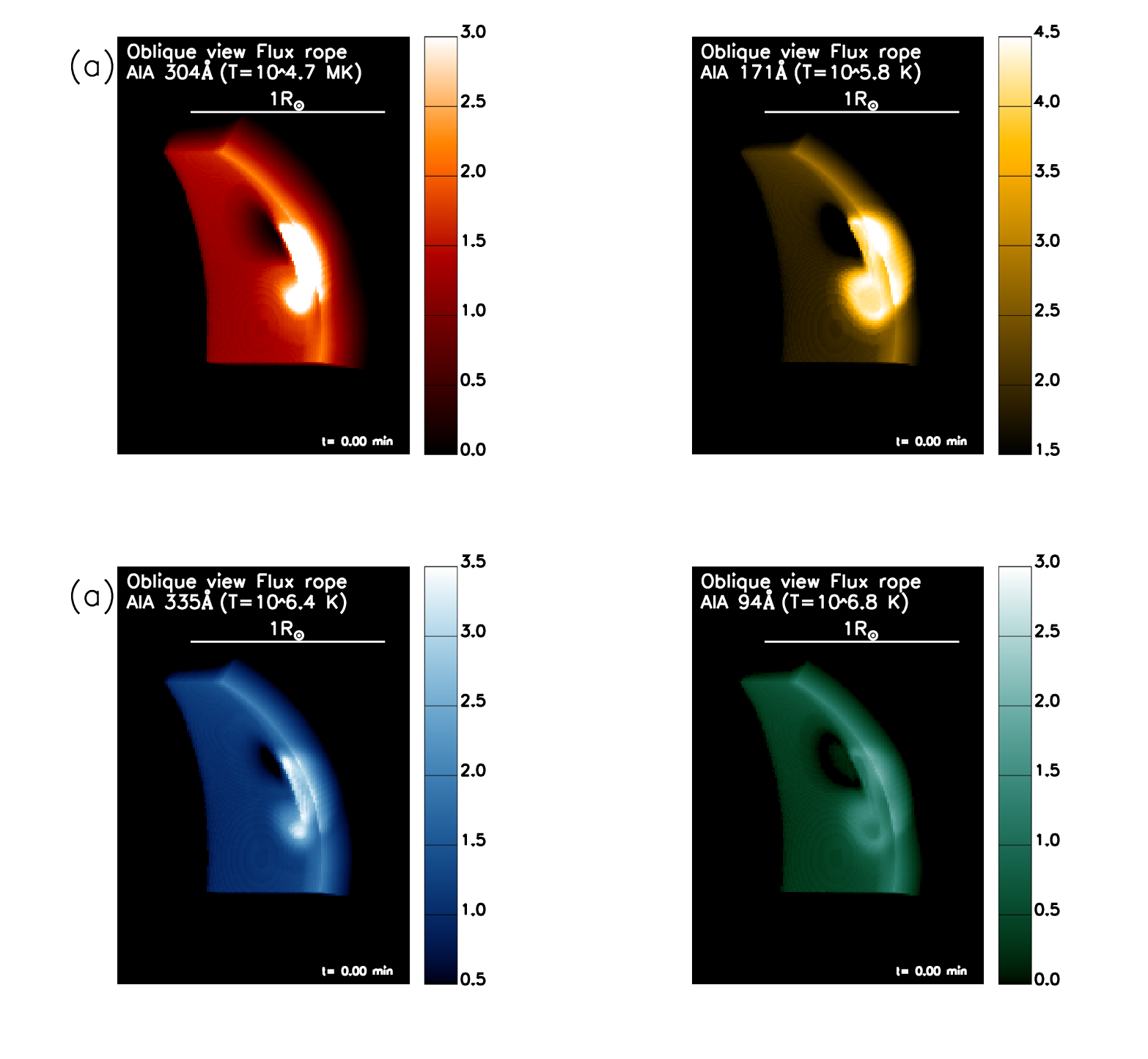}
\includegraphics[scale=0.49,clip,viewport=440 385 680 665]{fig910a.ps}
\includegraphics[scale=0.49,clip,viewport=045 385 315 665]{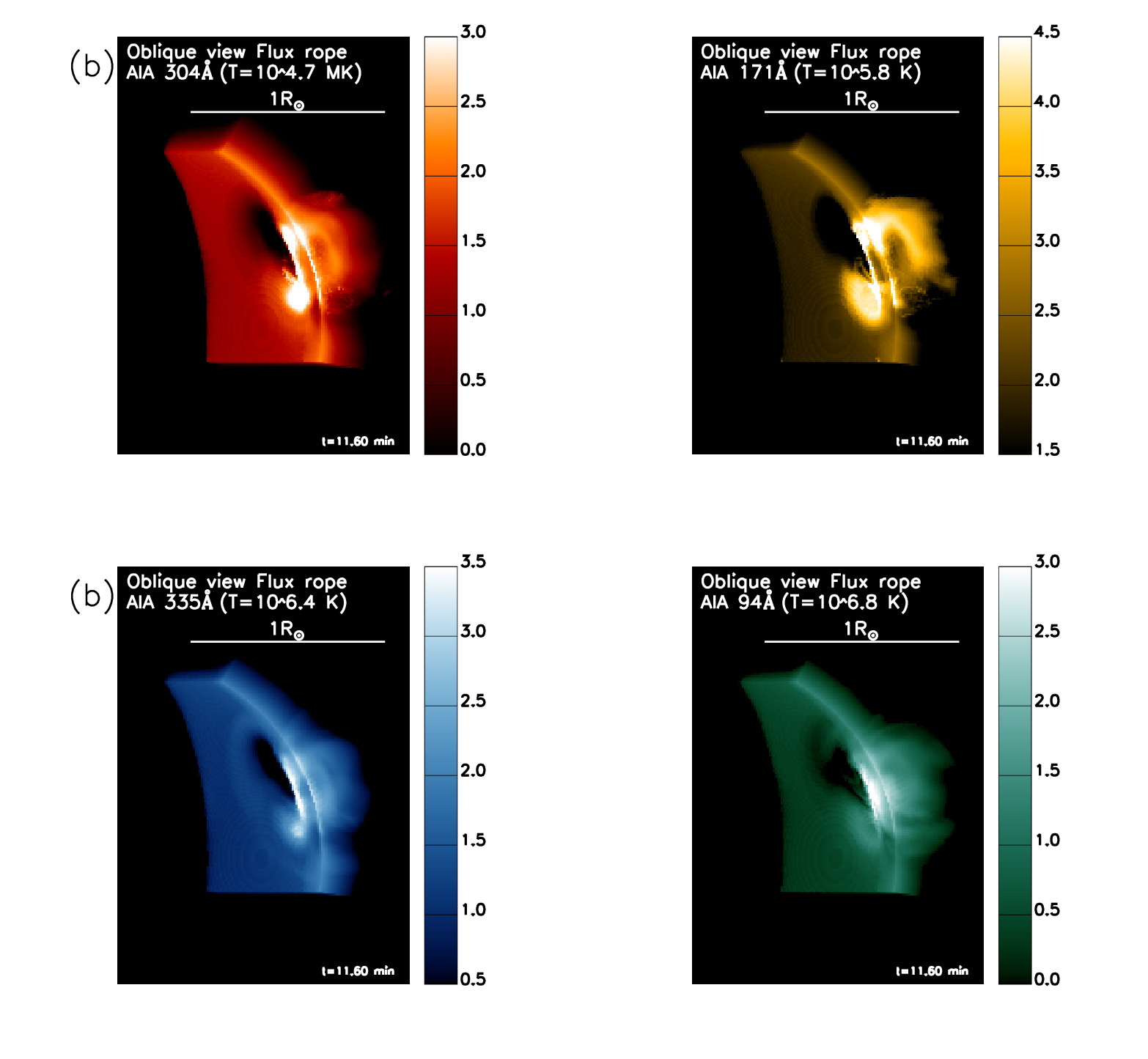}
\includegraphics[scale=0.49,clip,viewport=440 385 680 665]{fig910b.ps}
\includegraphics[scale=0.49,clip,viewport=045 385 315 665]{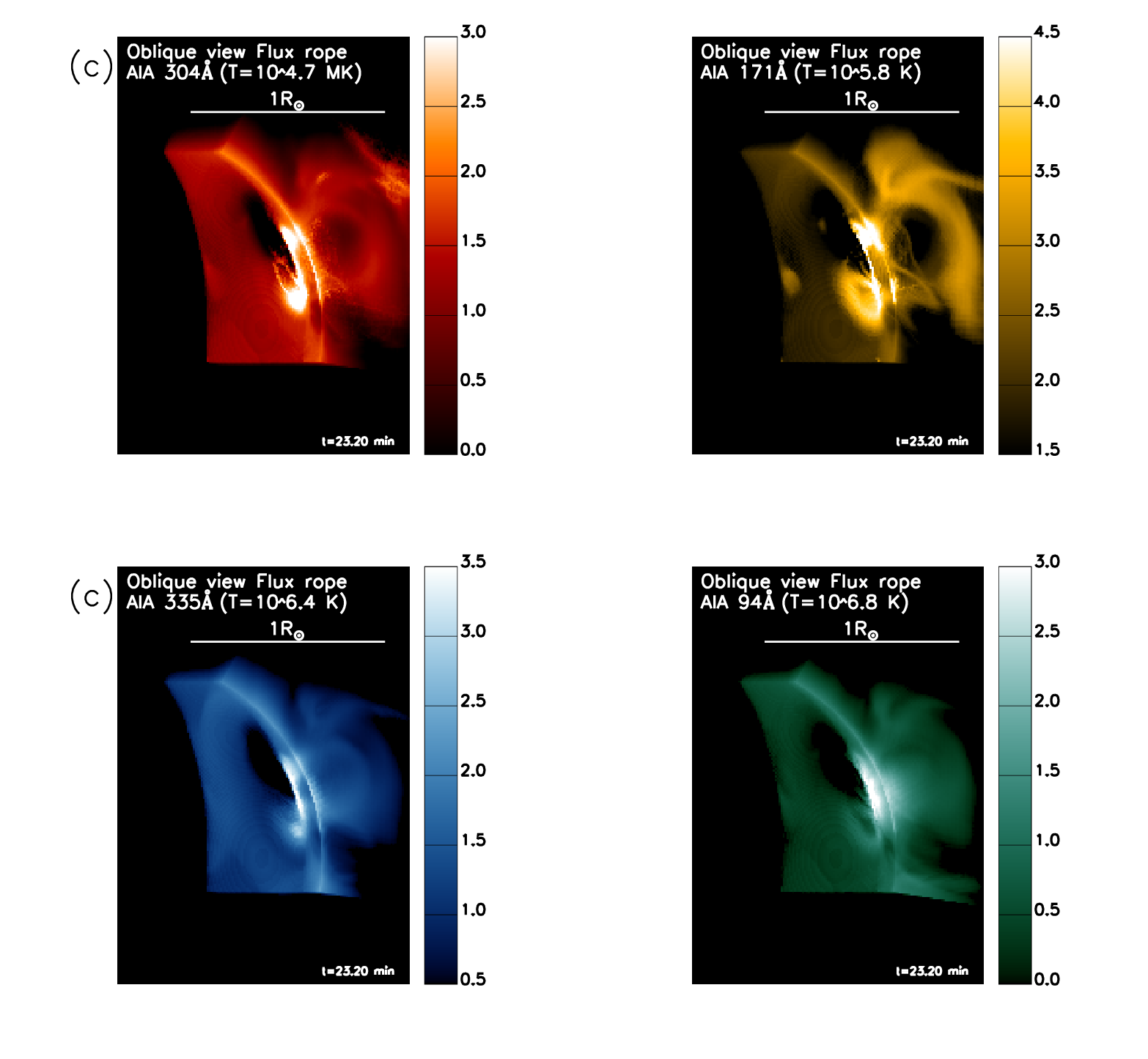}
\includegraphics[scale=0.49,clip,viewport=440 385 680 665]{fig910c.ps}
\caption{Synthesised observations of AIA in the $304\AA$ and $171\AA$ channels
at (a) $t=0$ $min$, (b) $t=11.20$ $min$, and (c) $t=23.20$ $min$.
All maps show the logarithm of synthesised DNS.
The temporal evolution is available in the online edition.}
\label{aiacold}
\end{figure}

Figure \ref{aiacold} shows the synthesised observations of the $304\AA$ and $171\AA$ channels
from the simulation at $t=0$, $11.60$, and $23.20$ $min$
where the flux rope initially lies on the solar surface
at $30^{\circ}$ from the plane of the sky in the direction of the observer
(a movie of Fig.\ref{aiacold} is available in the online edition).
This view point has no particular symmetry, and so is suitable to describe a general ejection.
The field of view used in Fig.\ref{aiacold} is $1.5\times2.15$ $R_{\odot}$ wide and represents 
the portion of the disk that can be viewed by AIA.
Figure \ref{aiacold} shows the logarithm of synthesised data numbers per second.

Initially the flux rope is clearly visible in the $304\AA$ and $171\AA$ channels  (Fig.\ref{aiacold}a).
The flux rope is slightly larger in the $171\AA$ channel,
as this channel picks out the external shell of the flux rope
as it is more sensitive to higher temperatures than the $304\AA$ channel.

At $t=11.60$ $min$ the ejection has already occurred and
the synthesised observations produce several features similar to the observations.
In both channels the eruption is clearly visible with an arc shaped structure
that is moving outwards.
In the $171\AA$ channel the structure is farther from the Sun, because of the different temperature sensitivity.
Similarly, the $171\AA$ channel shows a small void under the arc of the ejecting flux rope
which is less visible in the $304\AA$ channel.
In both channels it is possible to see some fragmentation of the structure at its sides.
In the $171\AA$ channel the strucuture seems to be more clearly anchored at its northern footpoint.
Comparing the position of the structures in Fig.\ref{aiacold}b
and the analysis of the position of the flux rope centre carried out in Fig.\ref{centerfrposition},
it is possible to infer that they both correspond to the ejected flux rope,
as the front has already propagated off the field of view.
It is important to note that
although at $t=11.60$ $min$ the flux rope temperature is far from the $10^{4.7}$ $K$
peak of the $304\AA$ channel, 
the ejection is still visible in this channel because of the high density of the flux rope plasma at $T=1$ $MK$,
a temperature at which the $304\AA$ channel still has a significant response (Fig.\ref{allfilters}).
Therefore, the $304\AA$ channel is observing hot and dense plasma rather than cold plasma here.
Similar but more expanded features are observed in the images at $t=23.20$ $min$ (Fig.\ref{aiacold}c).
In both channels the ejected flux rope has expanded and clearly looks like a
circular ejection with a void placed at the centre of the ejection.
The void is slightly more defined in the $171\AA$ channel and
the emission in the $304\AA$ channel is more diffuse.
In both channels,
the profile near the outer edge of the ejection
follows the arc shape with some features that
open up because of individual bundles of magnetic field lines that 
reconnected with the open magnetic field.
At the same time, it can be seen that the emission from the solar surface has been perturbed by the ejection.
In all of the images shown in Fig.\ref{aiacold},
we show the absolute data numbers per second (DNS) obtained in our study.
These values of DNS are comparable to within an order of magnitude
with actual off-limb observations from AIA during solar eruptions.

\subsection{Hot channels: $335\AA$ and $94\AA$}
\label{hotchannels}

\begin{figure}[!htcb]
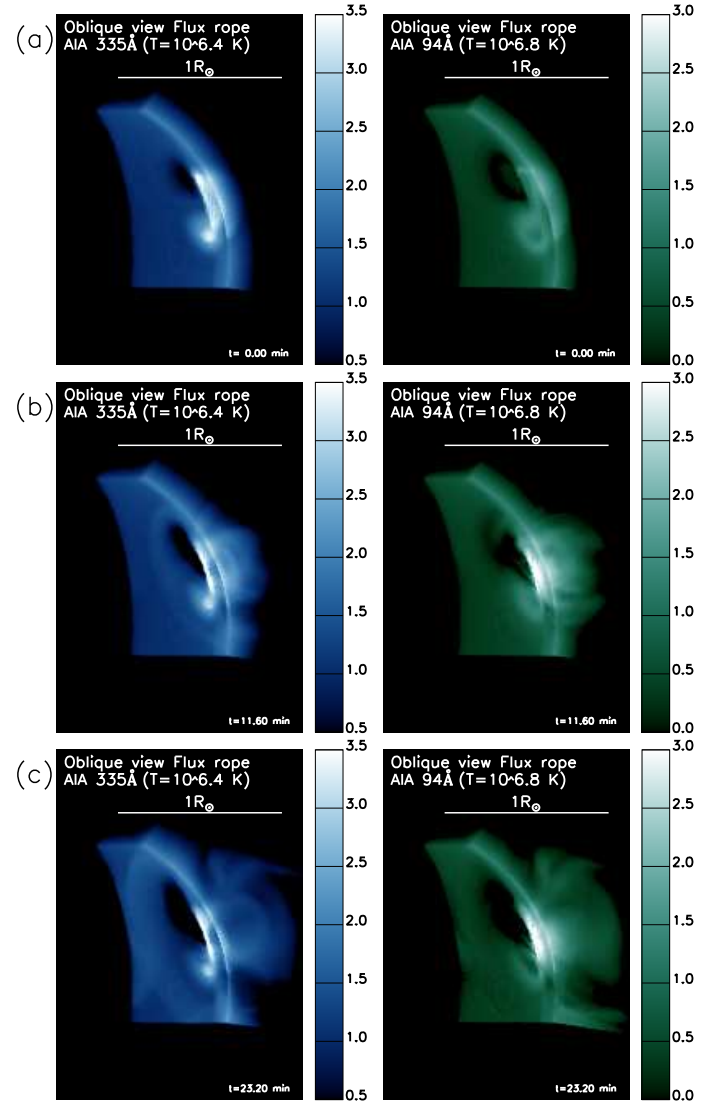

\centering
\includegraphics[scale=0.49,clip,viewport=045 045 315 325]{fig910a.ps}
\includegraphics[scale=0.49,clip,viewport=440 045 680 325]{fig910a.ps}
\includegraphics[scale=0.49,clip,viewport=045 045 315 325]{fig910b.ps}
\includegraphics[scale=0.49,clip,viewport=440 045 680 325]{fig910b.ps}
\includegraphics[scale=0.49,clip,viewport=045 045 315 325]{fig910c.ps}
\includegraphics[scale=0.49,clip,viewport=440 045 680 325]{fig910c.ps}
\caption{synthesised observations of AIA in the $335\AA$ and $94\AA$ channels
at (a) $t=0$ $min$, (b) $t=11.20$ $min$ and (c) $t=23.20$ $min$.
All maps show the logarithm of synthesised data numbers per second (DNS).
The temporal evolution is available in the on-line edition.}
\label{aiahot}
\end{figure}

The flux rope ejection appears significantly different when observed in
the $335\AA$ and $94\AA$ channels
when compared to the cold channels
(a movie of Fig.\ref{aiahot} is available in the online edition).
Initially, the flux rope is only visible in the $335\AA$ channel,
whereas the $94\AA$ channel only shows some brightness variation around the flux rope,
but no clear evidence of its existence (Fig.\ref{aiahot}a).

As the $335\AA$ and $94\AA$ channels are sensitive to temperatures hotter than those of the flux rope,
the arc-like erupting structure is not clearly visible during the ejection.
At $t=11.60$ $min$ (Fig.\ref{aiahot}b) the channels show a circular dome where an
arc is only slightly visible.
A brightening in the region behind the ejected flux rope can be seen because
it is heated to temperatures around $T\sim10^{6.5}$ relatively close to the peak temperature
of the $335\AA$ and $94\AA$ channels. 
In contrast, the region surrounding the flux rope ejection has lower temperatures
about $1$ $MK$ at this time (Fig.\ref{cutevol}) and it is very faint in these channels.
It is worth noting that in the $94\AA$ channel the region where the flux rope was originally placed
lights up after the ejection, as plasma at this location is heated because of the conversion of 
magnetic energy into thermal energy.

As the ejection evolves (Fig.\ref{aiahot}c), at $t=23.20$ $min$ we see that the dome has expanded in both 
the $335\AA$ and $94\AA$ channels, where a diffuse emission occurs over the whole region involved in the ejection.
It should be noted that the emission is still quite faint where both the $304\AA$ and $171\AA$ channels
show most of the ejected structures.
In the $335\AA$ and $94\AA$ channels the solar surface does not exhibit a significant variation due to the ejection.

\section{Discussion}
\label{discussion}

In the present paper we have synthesised AIA observations of a flux rope ejection.
The aim of this study is to show the observational signatures of the flux rope ejection model
with a simple tool
that can help verify the hypothesis upon which some models of solar eruptions are based.
This enables us to understand which physical processes are behind some of the observed features of CMEs.
In addition, we intend to address the relevance of the effects of thermal conduction and radiative losses
during flux rope ejections.

\subsection{Limits and assumptions of the present work}
\label{assumptionshere}

As we aim to reproduce actual observations of the Sun,
the assumptions made in our model
must be carefully accounted for in order to understand
the limitations of the study with
any sources of possible errors identified.
Here we address the effect of the 
assumptions made,
namely ionisation equilibrium, 
the magnetic configuration used,
and the lack of solar wind and transition region modelling.

The ionisation equilibrium assumption is reasonable for the initial condition, but
it becomes less realistic during the ejection.
\citet{Pagano2008} showed that ionisation equilibrium does not hold during a CME, at least for the propagating shock. 
However, the present work focuses especially on the ejected flux rope, where the ionisation times are significantly longer than in a shock region.
Figure \ref{cfrplots} shows that the flux rope undergoes an abrupt change in temperature only in the early phase of the ejection
after which the temperature slowly decreases.
We thus expect non-ionisation equilibrium to play a marginal role for the purposes of this study.
It was important to consider it, if we were to describe small-scale flaring emission
subsequent to the flux rope ejection initiation.
This additional aspect is, however, beyond the scope of the present paper.

Another feature that may be important is that
the initial magnetic configuration could be made more realistic,
as our system simply consists of two bipoles and a flux rope.
The actual solar corona is a much more complex environment, especially during solar maximum.
During maximum several CMEs per day may occur and
it is possible to have multiple active regions interacting or overlapping.
While this is the case, when the ejection occurs,
it is normally violent enough to displace the surrounding plasma and magnetic field.
Therefore the complexity around an active region should not play
a major role at least in the early stages of flux rope ejections.

In order to simulate the propagation of CMEs to larger distances from the Sun
we will have to include the effect of the solar wind.
The definitive coupling between CMEs and the solar wind occur at about $4$ $R_{\odot}$ \citep{Gopalswamy2000b},
thus it is not crucial in the spatial domain we consider,
however \citet{PomoellVainio2012} described the influence of solar wind in the thermodynamics of the plasma
during ejections.

Finally, if we wish to accurately describe the emission from the solar surface
in response to the ejection, we would need to include a transition region in our modelling.
This is particularly important for the cold channels whose response peaks at transition region temperatures.
While this is the case, as long as we focus on the off-limb emission, as we do in the present work,
the absence of a transition region emission should not significantly affect our conclusion.

In conclusion, it is worth to note that our synthesised observations are directly comparable
with actual ones, as the synthesised data number are close to actual counts (within an order of magntitude).
This shows that the density and temperature values reproduced by our MHD simulations are realistic,
at least for the region of corona involved.
In the $304\AA$ channel it is usual to observe $\sim10-100$ DN per second for off-limb
flux rope ejections and $\sim1000$ DNS for bright structures on the disk.
The $171\AA$ channel usually has about one order of magnitude larger DNS.
As far as the hot channels are concerned, observations count about $\sim1-50$ DNS for off-limb
flux rope ejections.
Of course, these values may vary when studying different events, but they are in reasonable agreement 
with our synthesis in Figs. \ref{aiacold} and Figs. \ref{aiahot}.
This proves that our model describes a realistic density and temperature distribution,
and the sum of assumptions is consistent with the aim of our work.

\subsection{Thermal conduction and radiative losses effects}

Our study also tries to shed some light on the importance of thermal conduction and radiative loss effects
when we study the ejection of a flux rope.
While, there is no doubt that thermal condution, radiative losses, and ohmic heating are all
physical processes actually taking place in the solar corona, the question is whether they are
effective during a flux rope ejection or CME.
A common drawback of the use of thermal conduction
is that it results in greater computational times.
A secondary effect of this is that with the 
increased number of timesteps needed to span a given physical time,
there are larger roundoff errors with an increase of the effect of numerical diffusion.
Therefore, it is worthwhile to question whether the use of these specific non-ideal terms is useful depending on the purpose of the investigation.

In the past, \citet{Reeves2010} made a significative attempt to model a flux rope ejection including 
non-ideal MHD terms and reported interesting findings on the energetics of the CME,
in particular with regard to how the ejection is initiated in a 2D domain.
Our work, instead, focuses on the propagation of the flux rope
and describes the effect of thermal conduction in three dimensions.
\citet{Lugaz2011} represents a key work in line with the attempt we have presented here
(although reproducing data from EIT).

In the study presented here, the non-ideal MHD terms play no role in the initiation of the eruption.
The eruption happens because of an unbalanced upward directed Lorentz force and, as explained in \citet{Pagano2013a},
the magnetic reconnection occuring behind the ejecting flux rope is a consequence of the flux rope's motion 
and it only reinforces the eruption, but it is not the main cause.

As the ejection proceeds, thermal conduction and radiative losses are effective where 
heat flows along magnetic field lines and the plasma radiates energy.
However, as shown in Fig.\ref{tempdiff} the effect of these terms in not dominant at the location of the ejection.
Whether we include or do not include non-ideal MHD terms,
the speed of the top and the centre of the flux rope (as in Fig.\ref{centerfrposition})
differs by only $3$ $km/s$.
From this point of view, it seems that the dynamics of the ejection are not affected by thermal conduction and radiative loss terms.
Thus, the consequences of heat flows and radiative losses are not fast and strong enough to 
affect the density, momentum, and magnetic field distribution which are primarily 
responsible for the dynamics of the ejection.

However, as we show in Fig.\ref{tempdiff}, thermal conduction
does have a significant impact on the temperature distribution.
Because of these effects there are differences of several $MK$s which cannot be neglected.
The plasma temperature is a parameter of crucial importance
in the solar corona for many reasons.
All the small-scale mechanisms (particle acceleration, plasma heating) are dependent on the temperature of the plasma
and they are often a important diagnostic tools with which to assess energetics and properties of large-scale events, 
such as solar eruptions and solar flares.

\section{Conclusions}
\label{conclusion}

In the present work, we synthesise EUV AIA observations of a flux rope ejection.
To do so we first perform a MHD simulation where a flux rope is ejected
as a consequence of it being initially out of equilibrium.
We then apply a simple model of EUV emission to compute the emission
from each plasma element in our simulation.
We finally take into account the response of each
AIA channel and integrate along the line of sight to obtain the synthesised observations.
Our work is a first step in this direction, but some issues can be improved in order
to have a better match between observations and models.

Our model shows a qualitative agreement
between the observations of CMEs as seen by AIA
and the simulations we have synthesised.
Many of the CMEs observed with AIA present 
a bright arcade rising in the $304\AA$ channel.
This feature is reproduced in our work (see also the online video related to Fig.\ref{aiacold}).
At the same time, we find that the south footpoint is less anchored than to
the north one (Fig.\ref{aiacold}b).
In principle, if this phenomenon were more pronounced,
leading to the complete reconnection of one footpoint with open flux, then
a different CME shape may be produced, such as a rising tail.
This can be ascribed to asymmetries around the erupting region,
where reconnection of the magnetic field is topologically favoured at one footpoint.

In our simulation, the flux rope is clearly visible from the initial stage
in the $304\AA$ and $171\AA$ channels.
In actual AIA observations, it is common to see a bright feature from where the CME originates.
While this is the case, the real magnetic configuration of the solar corona needs a more careful analysis before
we can conclude that there is a flux rope.
We can, however, claim that in both our simulation and in actual observations the ejection starts from a bright region in the $304\AA$ channel.
This indicates that to model the ejecting region
an initial temperature of around $10^{4}$ $K$ seems correct.

It should also be noted that the ejection is more evident and visible in the synthesised emission
than in the density maps we show in Fig.\ref{evolT2B6cdFR}.
The ejection of the flux rope does not lead to any local spike in density,
as the flux rope density always remains comparable to the surrounding plasma density.
In contrast, the displacement and expansion of the structures
in the $171\AA$ channel are evident thanks to the temperature effects
that sharply select the plasma we observe.
Whereas the emission in the $304\AA$ channel mostly originates from high density regions
at temperatures about the tail of the filter response, as we explained in Sect. \ref{coldchannels}.
In our study, the $171\AA$ and $304\AA$ channel results are the most appropriate to follow 
the flux rope propagation, because of the flux rope temperature.
In contrast, the hot channels at $335\AA$ and $94\AA$ are suitable to highlight the
temperature increase of the plasma during the CME.
In our simulation, the heating occurs at the front of the ejection due to compression (not visible in Fig.\ref{aiahot}b),
and behind the propagating flux rope where
magnetic energy is converted into thermal energy which results in a diffuse emission (Fig.\ref{aiahot}b-c).
This process is very similar to a reconnection event at the current sheet as prescribed
by the standard flare model and also observed in the $131\AA$ channel by \citet{Cheng2011}
from a different line of sight than that considered within this study.

In future, we plan to extend our coupling technique between the GNLFFF model
and the MHD simulation to include the use of real magnetograms 
representing the full Sun
as an initial condition in our MHD simulations
\citep[such as used in][]{Yeates2010}.
This will reproduce more complex and realistic patterns of the solar corona
that result in flux rope ejections.
We will also focus on modelling specific events using 
idealised initial magnetic configurations inspired by observed magnetic structures prior to ejections
and by real magnetograms prior to actual ejections.
Even though the above improvements are required,
the current status of the modelling technique is sufficient
to reproduce many of the main features presently found in observations and
allow a thorough understanding and the synthetis of AIA images.

Another aspect of the work presented here
is that we are not restricted to constructing synthetic AIA observations.
The same technique may be applied to the EUV channels observed by the EUI instrument on board Solar Orbiter.
Thus, the development of this technique provides a platform for the prediction of features that may be observed by Solar Orbiter.

\begin{acknowledgements}
We acknowledge the use of the open source (gitorious.org/amrvac) MPI-AMRVAC software, relying on coding efforts from C. Xia, O. Porth, R. Keppens. 
DHM would like to thank STFC, the Leverhulme Trust and the European Commission's Seventh Framework Programme
(FP7/2007-2013)  for their financial support.  PP would like to thank the European Commission's Seventh Framework Programme
(FP7/2007-2013) under grant agreement SWIFF (project 263340, www.swiff.eu) and STFC for  financial support.
These results were obtained in the framework of the projects
GOA/2009-009 (KU Leuven), G.0729.11 (FWO-Vlaanderen) and C~90347 (ESA Prodex 9).
The research leading to these results has also received funding from the European Commission's Seventh Framework Programme (FP7/2007-2013) under the grant agreements
SOLSPANET (project n 269299, www.solspanet.eu),
SPACECAST (project n 262468, fp7-spacecast.eu),
eHeroes (project n 284461, www.eheroes.eu).
The computational work for this paper was carried out on the joint STFC and SFC (SRIF) funded cluster at the University of St Andrews (Scotland, UK).
\end{acknowledgements}

\bibliographystyle{aa}

\end{document}